%% file: Loa_NT-HP-Review.tex
\documentclass[a4paper,a4wide,11pt]{article}

\usepackage[]{graphicx}           

\usepackage[nosort]{cite}

\usepackage{textcomp}              
\usepackage[T1]{fontenc}
\usepackage[ansinew]{inputenc}
\usepackage{textscripts}           

\usepackage{amsmath}
\usepackage{amssymb}

\usepackage{3parttable}
\usepackage{booktabs}
\usepackage{dcolumn}
\newcolumntype{d}[1]{D{'}{}{#1}}

   \newcommand{\WN}{\mbox{cm\sups{$-1$}}}

   \renewcommand{\d}{\ensuremath{\mathrm{d}}}

   \renewcommand{\vec}[1]{\ensuremath{\boldsymbol{#1}}}%
   \newcommand{\mum}{\textmu m}
   \newcommand{\degreeC}{\mbox{\textdegree C}}
   \newcommand{\To}{\mbox{--}}
   \newcommand{\etal}{\textit{et~al.}}
  \newcommand{\doublespace}{\renewcommand{\baselinestretch}{1.1}}
  \newcommand{\singlespace}{\renewcommand{\baselinestretch}{1.00}}
  \newcommand{\ccol}[1]{\multicolumn{1}{c}{#1}}
  \newcommand{\mc}[3]{\multicolumn{#1}{#2}{#3}}

\newcommand{\ns}{\normalsize}

\begin{document}

\title{\textbf{Raman spectroscopy on carbon nanotubes\\ at high pressure}}
\author{I. Loa\\
    \ns Max-Planck-Institut für Festkörperforschung,
    D-70569 Stuttgart, Germany}
\date{}

\maketitle

\noindent\rule{\linewidth}{0.75pt}%
\begin{quote}
\noindent\textbf{Abstract}

\noindent Raman spectroscopy has been the most extensively employed method
to study carbon nanotubes at high pressures. This review covers
reversible pressure-induced changes of the lattice dynamics and
structure of single- and multi-wall carbon nanotubes as well as
irreversible transformations induced by high pressures. The interplay
of covalent and van-der-Waals bonding in single-wall nanotube bundles
and a structural distortion near 2~GPa are discussed in detail.
Attempts of transforming carbon nanotubes into diamond and other
``superhard'' phases are reviewed critically.

\end{quote}
\noindent\rule{\linewidth}{0.75pt}%

\vfill

\begin{quote}
    \sffamily
    Ingo Loa\\
    Max-Planck-Institut für Festkörperforschung\\
    Heisenbergstr.\ 1\\
    D-70569 Stuttgart\\
    Germany\\[1ex]
    phone +49.711.689--1469; fax --1444\\
    E-mail: I.Loa@fkf.mpg.de\\
\end{quote}

\vfill
\centerline{\textsf{To appear in J. Raman Spectroscopy (July/August
2003): Special Issue on >>Extreme Conditions<<.}}


\clearpage

\tableofcontents

\newpage

\doublespace\normalsize

\section{Introduction}

Single-wall carbon nanotubes (SWNTs) are tubular macromolecules with
typical diameters on the order of 1~nm and lengths up to several \mum\
that are made purely from carbon. As a consequence of their small
diameter and very large length-to-diameter ratio, \emph{isolated}
nanotubes represent nearly ideal realizations of a one-dimensional
system, in particular with regard to their electronic properties. Solid
SWNT material produced by the common growth techniques, however,
usually consists of tangled bundles containing up to a few hundred
aligned nanotubes. Mechanically, carbon nanotubes combine high
stiffness and tensile strength with low density. Their fascinating
structure and special characteristics as well as the prospect of
numerous technical applications has led to intense research on this
material over the last decade. Part of the potential applications
\cite{AZ01} -- including electron field emission \cite{FCFT99},
field-emission displays \cite{WSLD98,CCKK99}, hydrogen storage
\cite{LFLC99}, single electron transistors \cite{TVD98}, gas sensors
\cite{KFZC00}, and nanomechanical devices \cite{KL99a} -- have been
realized during recent years. The discovery of the single-wall carbon
nanotubes in 1993 \cite{II93,BKVG93} was preceded by the observation of
multi-wall nanotubes (MWNTs) two years earlier \cite{Iij91}. The latter
consist of a coaxial series of carbon tubes nested one inside the
other.

Application of high pressures is the ideal tool to tune continuously the
bonding properties of a solid. The structural variations affect virtually
all material properties. In the context of carbon nanotubes, the
pressure-induced changes of the vibrational characteristics are of
particular interest. They yield, although indirectly, important
information on the bonding properties, especially with regard to the
interplay of covalent and van-der-Waals (vdW) bonding. In addition,
structural transformations manifest themselves in changes of the lattice
dynamics. The vibrational properties of nanotubes in a high-pressure
diamond anvil cell (DAC) can readily be investigated by Raman spectroscopy
which has played an important role in the field of carbon nanotubes for a
long time.

This review is organized as follows. Section~\ref{sec:background}
summarizes some basic background information on the properties and
synthesis of carbon nanotubes. Reversible pressure-induced changes of the
lattice dynamics and structure of single-wall and multi-wall nanotubes
will be discussed in sections~\ref{sec:SWNT-lattice-dynamics} and
\ref{sec:MWNT-lattice-dynamics}, respectively.
Section~\ref{sec:structural-transformations} addresses the question of
irreversible pressure-induced structural transformations, especially
towards hard phases such as diamond. Finally, perspectives on possible
future developments are briefly presented in the concluding
section~\ref{sec:summary}.

\clearpage

\section{Background}
\label{sec:background}

Comprehensive treatments of the synthesis, characterization, properties
and applications of carbon nanotubes were presented in several books
and review articles, see e.~g.\ Refs.~\cite{DDE96,SDD98,DDA01,DE00}.
Here, we shall summarize briefly some basic information to make this
short review reasonably self-contained.

\subsection{Basic Properties of Carbon Nanotubes}

\begin{figure}[t]
     \centering
     \includegraphics[width=0.8\hsize]{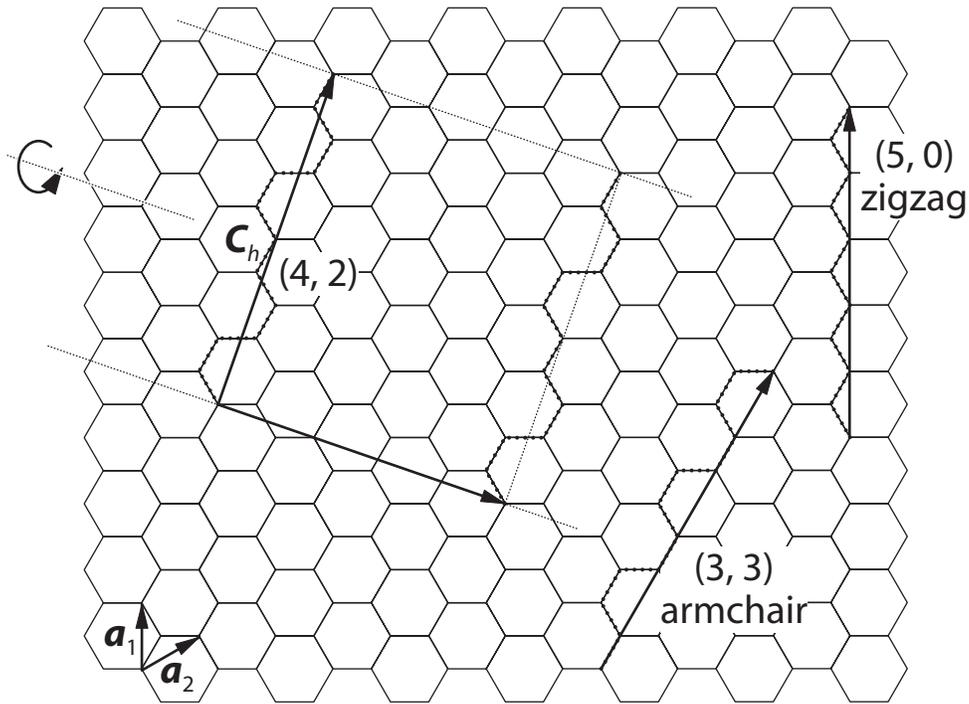}
     \caption{Rolling-up of a graphene sheet into a nanotube. The geometry
              of a single-wall nanotube is defined by the roll-up vector
              $(n, m)$ or, equivalently, the chiral vector $\vec C_h =
              n\vec{a}_1 + m\vec{a}_2$. The chiral vectors of
              a zigzag and an armchair tube are also shown.}
     \label{fig:graphene-roll-up}
\end{figure}

Single-wall carbon nanotubes can be thought of as being formed by
rolling a graphene sheet into a cylinder
(Fig.~\ref{fig:graphene-roll-up}). Depending on how the sides of the
sheets are joined, nanotubes of different diameter and helicity are
obtained. All possible structures are uniquely identified by a pair of
integer numbers, the \emph{roll-up vector}\/ $(n, m)$. It defines the
chirality vector $\vec C_h = n\vec{a}_1 + m\vec{a}_2$ that connects two
symmetry-equivalent atoms of the 2D graphene sheet
(Fig.~\ref{fig:graphene-roll-up}). After rolling the graphene sheet
into a tube, the chiral vector becomes a circumferential line of the
tube. SWNTs with a roll-up vector of the form $(n, 0)$ are called
\emph{zigzag}\/ tubes. \emph{Armchair} tubes are characterized by a
roll-up vector $(n,n)$. Both types are \emph{achiral}\/ tubes; all
other tubes are referred to as \emph{chiral}\/ nanotubes. The diameter
of a carbon nanotube is given by
\begin{equation}\label{eq:NT-diameter}
d_{nm} = |\vec C_h|/\pi =
\sqrt{3}a_{\text{C--C}}(m^2 + nm + n^2)^{1/2}/\pi,
\end{equation}
where $a_{\text{C--C}}$ denotes the carbon--carbon bond length
(1.42~Å). Single-wall nanotubes grown by the common arc-discharge or
laser-ablation methods \cite{SDD98,Dai01} usually exhibit a narrow
distribution of diameters. A typical average diameter of 1.4~nm is
close to that of a $(10,10)$ nanotube. SWNTs do not generally exist as
individual tubes. They rather form bundles of up to a few hundred tubes
in a regular hexagonal lattice \cite{TLND96,II93,BKVG93}.

The electronic properties of SWNTs depend on their structure. There are
metallic, semi-metallic, and semiconducting SWNTs with electronic band
gaps up to $\sim$1~eV. Being derived from graphene sheets, carbon
nanotubes show predominantly $sp^2$-type bonding of the carbon atoms.
Deviation from the planar bonding introduces, however, some $sp^3$-type
hybridization. In view of the strong covalent bonding, carbon nanotubes
have been expected to exhibit high stiffness and tensile strength.
Measurements of mechanical quantities are difficult to perform on
individual tubes. Nonetheless, such experiments were carried out and large
values of Young's modulus on the order of 1~TPa were reported
\cite{KDEY98,PWUH99,YA01}.

\begin{figure}[t]
     \centering
     \includegraphics[width=0.8\hsize]{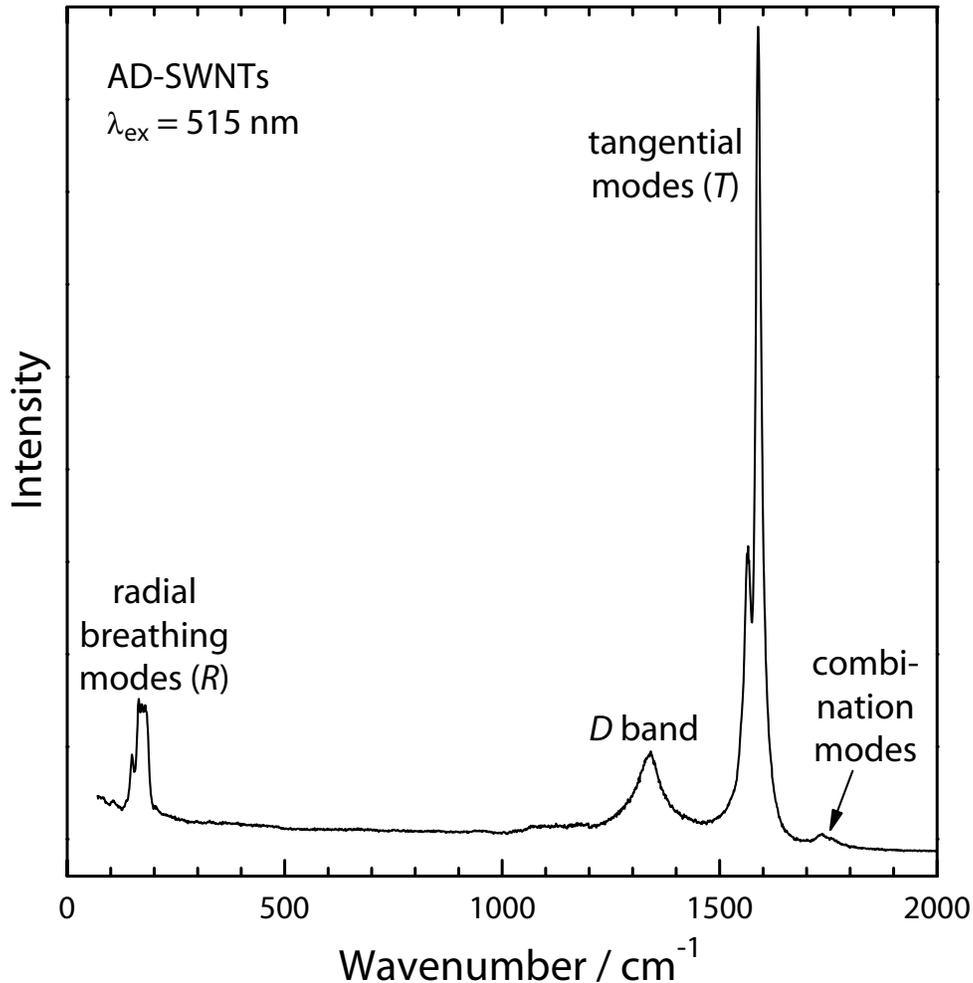}
     \caption{Raman spectrum of arc-discharge-grown SWNTs at ambient
     conditions ($\lambda_\text{ex}=515$~nm).}
     \label{fig:SWNT-Raman-ambient}
\end{figure}

Raman spectroscopy has become a widespread tool for the investigation as
well as the routine characterization of carbon nanotubes \cite{SDD98}. It
yields information on the vibrational and, indirectly, on the structural
properties. Raman spectra of SWNTs typically show intense peaks in the
spectral ranges of 160--200 and 1500--1600~\WN
(Fig.~\ref{fig:SWNT-Raman-ambient}). The first assignment of these
features was given by Rao \etal\ \cite{RRBC97}. The low-energy peaks are
attributed to a \emph{radial breathing mode} ($R$ band), where all C atoms
are subject to an in-phase radial displacement. This mode is rather
strongly diameter-dependent and was therefore proposed as a means to
determine the SWNT diameter \cite{BASR98}. The higher-energy peaks are
attributed to modes with neighboring C atoms vibrating out-of-phase
parallel to the surface of the cylinder (\emph{tangential modes}, $T$
band) \cite{RRBC97} which are related to the $E_{2g}(2)$ phonon at
$\sim$1580~\WN\ in graphite \cite{DD82,TK70,HBS89,PHKL98}. In achiral
tubes, one can distinguish two tangential modes with atomic displacements
parallel and perpendicular to the tube axis, respectively. Besides these
Raman-active vibrations, there exists a defect-related mode near 1350~\WN\
that is known from graphite \cite{NS79}. Overtone and combination modes of
the above excitations have also been observed.

Multi-wall carbon nanotubes usually consist of several tens of
concentric shells. The spacing between the sheets equals approximately
the interlayer distance in graphite (3.4~Å). Like in the case of the
SWNTs, bending of the graphene sheets has little influence of the
structural and electronic properties. This applies all the more to
MWNTs because of their larger diameter, i.~e., smaller curvature of the
layers. Besides the geometry, the most important difference between
MWNTs and graphite is the confinement of the charge carriers to the
tubular structure of the nanotubes. Many physical properties of MWNTs
are quite similar to those of graphite.

\subsection{Synthesis of Carbon Nanotubes}
Until recently, two methods have dominated the production of carbon
nanotubes: \emph{arc-discharge} (AD) growth and synthesis by
\emph{pulsed-laser vaporation} (PLV) \cite{EA92,BKVG93,TLND96,Dai01}.
Both methods are based on the evaporation of carbon from a solid
source. In the arc-discharge method, a He plasma is created between two
opposing carbon electrodes. Carbon is evaporated at temperatures of
3000--4000~K. It condenses at lower-temperature areas of the cathode
and forms nanotubes and other carbon  particles, depending on the
growth conditions. For the growth of SWNTs yttrium and/or transition
metals need to be used as a catalyst. A typical mixture for the anode
contains 1 at.~\% of Y and 4.2~at.~\% of Ni besides the carbon
(graphite) \cite{JMBL97}. In the PLV process, evaporation of carbon is
achieved by laser ablation from a carbon target in a furnace at
$\sim$1500~K \cite{TLND96}. The characteristics, in particular the
diameter distribution and the yield of the nanotubes can be controlled
by the choice of catalysts and growth conditions (inert gas pressure,
discharge current, \ldots).

Removal of the metal catalyst particles and by-products such as
fullerenes, graphitic polyhedrons, and amorphous carbon can be achieved by
a purification process. It involves refluxing the as-grown material in a
nitric acid solution, resuspending the nanotubes in pH-10 water with a
surfactant, and filtration \cite{LRDH98}, possibly followed by a
heat-treatment in vacuum to remove residues of the chemical treatment.
This method produces a free-standing mat of tangled SWNT ropes, so-called
\emph{bucky paper}.

As an alternative, chemical vapor deposition (CVD) has been employed for
the growth of carbon nanotubes \cite{Dai01}. It had been used for the
production of carbon fibers and filaments before the advent of carbon
nanotubes. While it has been challenging to produce structurally perfect
nanotubes by this method, it offers the possibility to grow ordered and
aligned nanotube structures \cite{DKZF99,Dai01}. Furthermore, nanotube
growth at specific locations on a surface can be achieved by employing
contact printing or lithographic techniques to control the deposition of
the catalyst on a substrate \cite{Dai01}.

\clearpage

\section{Lattice Dynamics and Structure of SWNTs}
\label{sec:SWNT-lattice-dynamics}

The vibrational properties of single-wall carbon nanotubes under
hydrostatic pressure have been investigated in numerous Raman scattering
studies. In the following two sections, results on the radial mode and the
tangential modes will be reviewed, respectively. In either case, there are
indications of a structural distortion or transition near 2~GPa that will
discussed in section~\ref{sec:anomaly-2GPa}.

\subsection{Radial Breathing Mode}

\begin{figure}[t]
     \centering
     \includegraphics[height=9cm]{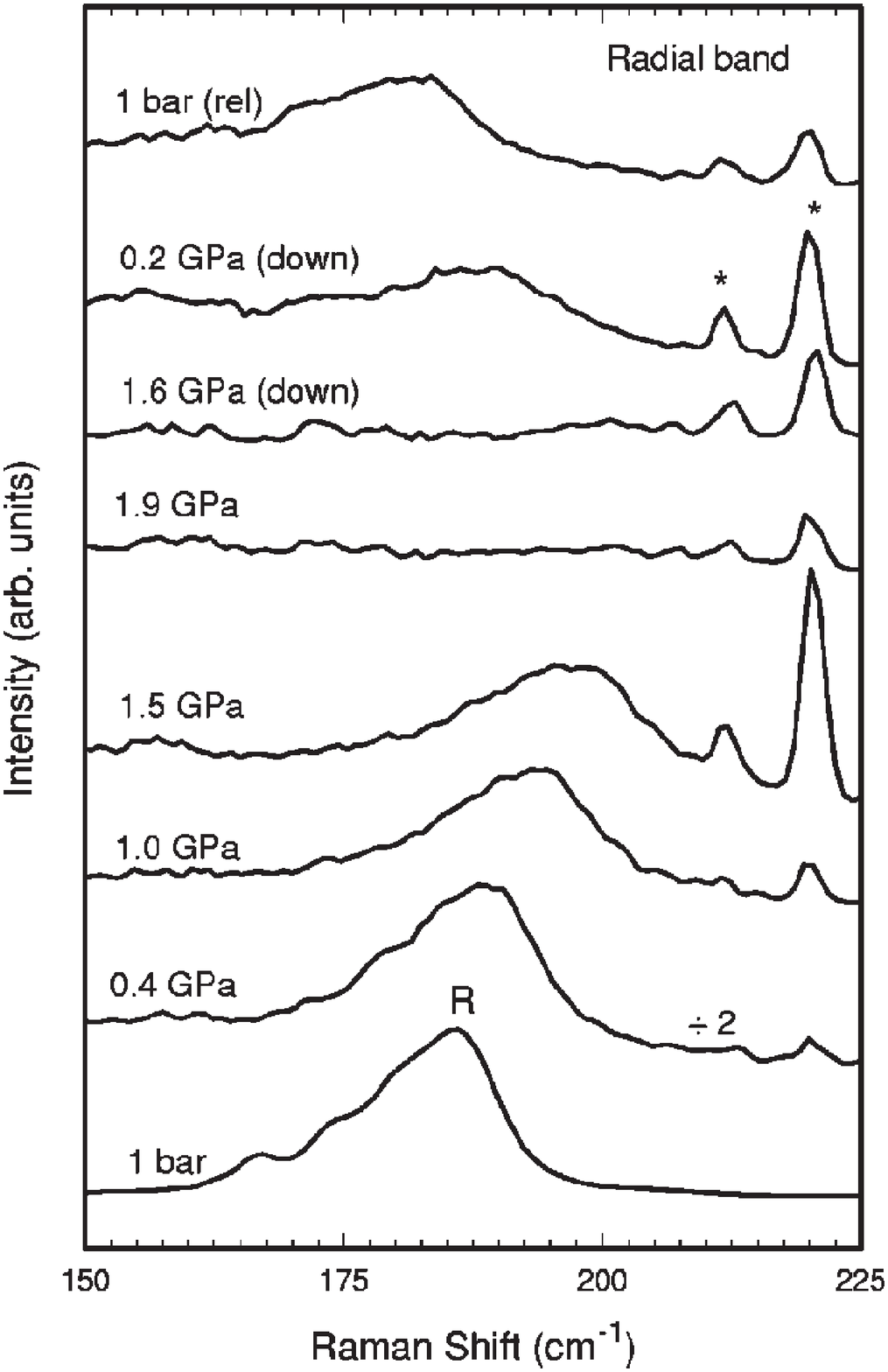}
     \hspace{2em}
     \includegraphics[height=7cm]{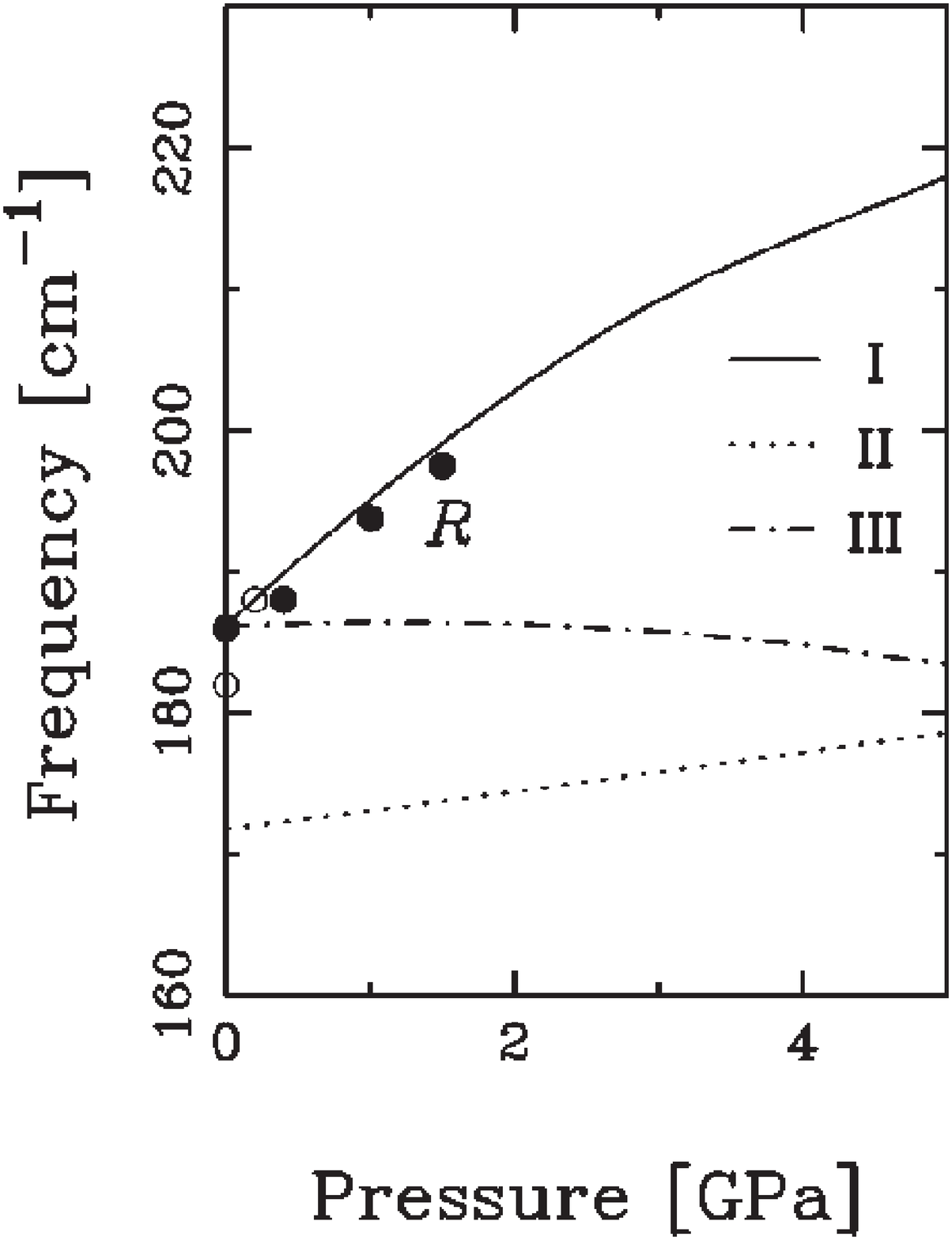}
     \caption{(left) Raman spectra of PLV-SWNT bundles in
     the region of the radial breathing mode as a function of pressure
     ($T=300$~K, excitation wavelength $\lambda_{ex}=515$~nm). The $R$
     band disappears beyond 1.5~GPa. (right) Pressure dependence of the
     radial mode wavenumber. Solid (open) symbols refer to data measured
     during the upward (downward) cycle of pressure. The lines show
     results of GTBMD calculations (models I--III in Ref.~\cite{VRRM99})
     for a $(9,9)$ nanotube bundle. Reproduced from Ref.~\cite{VRRM99}.}
     \label{fig:RBM-Specs-VRRM99-1a-3a}
\end{figure}

\paragraph{Basic Results}
The first study of the effect of hydrostatic pressure on the Raman
spectrum of SWNTs was published by Venkateswaran \etal\ \cite{VRRM99}.
Figure~\ref{fig:RBM-Specs-VRRM99-1a-3a} shows Raman spectra of as-prepared
PLV-SWNT bundles in the spectral region of the $R$~band \cite{VRRM99}. The
common 4:1 methanol/ethanol mixture was used as the pressure transmitting
medium in the diamond anvil cell. The $R$ band shifts towards higher
wavenumbers with increasing pressure at rate of $7 \pm 1$~\WN/GPa. The $R$
band decreases in intensity as the pressure is increased, and it vanishes
beyond 1.5~GPa in contrast to the higher-energy tangential modes (section
\ref{sec:TM-Raman}). Upon pressure reduction from 5.2~GPa, the $R$ band
reappears around 1.5~GPa. After the pressure cycle, the $R$ mode recovered
only part of its initial intensity.

In place of the common methanol/ethanol pressure medium, Obratzsova
\etal\ \cite{OLSK99} used condensed nitrogen as a pressure transmitter,
that is fluid up to 2.5~GPa (at ambient temperature) and provides
quasi-hydrostatic conditions to several tens of GPa. Here, the $R$ band
of PLV bucky paper was observed to shift at a rate of $\sim$10~\WN/GPa.
This is larger than the shift observed by Venkateswaran \etal\
\cite{VRRM99}, but consistent with results of Peters \etal\
\cite{PMLK00}, Thomsen \etal\ \cite{TRGJ99,TRJL99}, and Sood \etal\
\cite{STMS99,TSMS00}. Teredesai
\etal\ \cite{TSSK01} employed water as a pressure medium, that solidifies
near 1~GPa (at 300~K), and found for the $R$ band essentially the same
pressure dependence as in the case of a 16:3:1 mixture of
methanol/ethanol/water. Altogether, pressure shifts of the $R$ band
wavenumber in the range 7--10~\WN/GPa have been reported, with no clear
correlation between the shift rate and the sample type or experimental
conditions. Table~\ref{tab:Raman1} summarizes the numerical results of
all the studies cited here and in the next section.

\begin{table}
  \singlespace
\begin{threeparttable}
  \setlength{\defaultaddspace}{0.5ex}

  \caption{Compilation of Raman scattering and theoretical results on
  carbon nanotube phonon wavenumbers $\omega_0$ and their respective
  pressure dependences $\d \omega/\d P$.
  The experimental $T$-band data refer to the most intense component of
  the $T$-band. Grouped together are experiments that appear to have been
  performed on samples of the same origin. A 4:1 methanol/ethanol mixture
  (M/E) was used as a pressure medium, unless noted otherwise (H$_2$O,
  N$_2$, He, M/E/W \{16:3:1 methanol/ethanol/water\}).}
  \label{tab:Raman1}

  \scriptsize
\input{Loa-Raman-table.inc}
\end{threeparttable}
\end{table}

\begin{figure}[t]
     \centering
     \includegraphics[width=0.8\hsize]{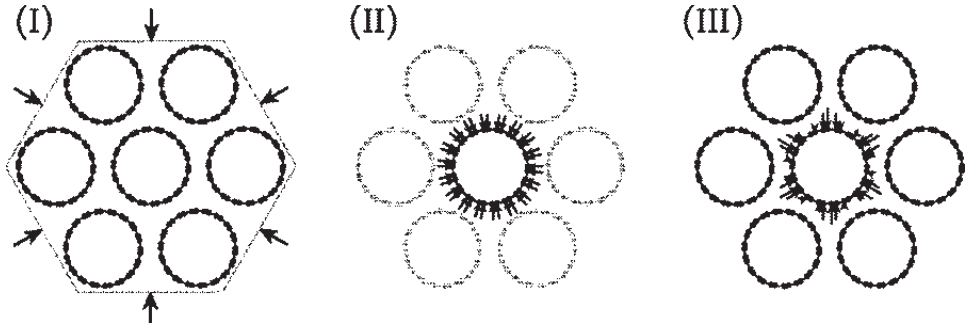}
     \caption{Cross-sectional view of a SWNT bundle showing schematically
          the forces on the bundle and the individual tubes in the triangular
          lattice when hydrostatic pressure is applied. In model I, the entire
          bundle is subjected to an external compression. In model II, the
          individual tubes are each compressed symmetrically, and
          intertubule coupling is ignored. The absence of vdW interactions
          between the tubes is shown schematically by the lightly shaded
          tubes surrounding the center tube. In model III, the pressure
          medium is allowed to penetrate into the interstitial channels
          between tubes and thereby exerts a $\sin^2 3 \Theta$ force on the
          tubes. Models I and III include intertubule vdW coupling.
          Reproduced from Ref.~\cite{VRRM99}.}
     \label{fig:TM-VRRM99-2}
\end{figure}

\paragraph{Lattice Dynamical Calculations}
To aid the interpretation of their experimental results, Venkateswaran
\etal\ \cite{VRRM99} performed calculations of the radial and tangential
modes frequencies of a $(9,9)$ SWNT bundle, using a generalized
tight-binding molecular dynamics (GTBMD) scheme. Going beyond the original
formulation of this method \cite{MRS96}, they treated the
van-der-Waals-type (vdW) intertubule interaction by the addition of a
Lenard-Jones-type potential to the GTBMD. Three models of the interaction
of the pressure medium with the nanotube bundles were considered as shown
in Fig.~\ref{fig:TM-VRRM99-2}. Comparison of the theoretical and
experimental results in Fig.~\ref{fig:RBM-Specs-VRRM99-1a-3a} shows that
only their model I, where the pressure medium does not penetrate the
bundles, is in agreement with the experiment. Comparison between model I
and model II (without vdW forces) demonstrates the importance of the
initially weak intertubule interaction for the experimentally observed
pressure dependence of the radial mode frequency. None of the pressure
media alcohol, water, and nitrogen that have been employed in the various
studies appears to penetrate into the interstitial channels (diameter
$\sim$2.6~Å \cite{VRRM99}) between the bundled carbon nanotubes.

\paragraph{Intertubule Interaction}
It was noticed \cite{VRRM99} that inclusion of the intertubule vdW
interaction in the calculation leads to an up-shift of the
zero-pressure $R$ mode wavenumber by $\sim$14~\WN\ ($\sim$8\%). Further
tight-binding calculations by Kahn and Lu \cite{KL99b} as well as
Henrard \etal\ \cite{HHBR99,HPR01} predicted a 6--24~\WN\ increase in
wavenumber for the $R$ modes in bundled nanotubes with diameters of
10--16~Å compared to those of the respective isolated tubes. These are
important results as the $R$ mode wavenumber $\omega_R$ was proposed
\cite{BASR98,KKK98} as a measure for the nanotube diameter $d_t$
(Eq.~\ref{eq:NT-diameter}) via the relation $\omega_R = 223.75 (\WN
\times \mbox{nm}) / d_t$. If one neglects the small dependence
\cite{HPR01} of the vdW-related wavenumber shift on nanotube geometry
for the range of tube diameters of experimental interest, this leads to
a modified relation between $R$ mode wavenumber and tube diameter,
$\omega_R
\approx 224 (\WN \times \mbox{nm}) / d_t  + \Delta_\text{vdW}$, where
$\Delta_\text{vdW}$ is on the order of 15~\WN. However, recent
experiments \cite{RCRS01} on very small bundles containing only a few
nanotubes, so-called solubilized tubes (s-SWNT \cite{CHHC98,CRLI01}),
have evidenced a small up-shift of the $R$ mode Raman peak in the
\emph{debundled} tubes compared to the bundled ones. The apparent
contradiction to the above lattice dynamical predictions was resolved
by also taking into account the effect of intertube interaction on the
electronic density of states as well as the known diameter-selective
resonance effects in the Raman scattering process
\cite{RRBC97,RS97,MKHK00}. Altogether, the intertube interaction has
been estimated to lead to a down-shift of the $R$ mode \emph{Raman
peak}\/ of $\sim$4~\WN\ for metallic tubes and $\sim$40~\WN\ for
semiconducting tubes, despite the up-shift of the $R$ mode
\emph{vibrational frequency}\/ on the order of 15~\WN. It is unclear at
present to what extent the pressure-induced changes of the electronic
structure affect the shift of the $R$ mode Raman peak under pressure.

The effect of intertubule interaction was further studied by Schlecht
\etal\ \cite{VBSR01,SVRC02p} in a high-pressure experiment on the
above-mentioned solubilized nanotubes. From a height-profile analysis
of scanning force microscopy images \cite{CHHC98,CRLI01}, these s-SWNTs
were deduced to contain nanotubes mostly in small bundles of 3--7
tubes, but also some isolated tubes. Although the ambient-pressure $R$
mode energy is affected by the reduction in bundle size \cite{RCRS01},
they observed, surprisingly, a shift rate of 8.4~\WN/GPa which equals
the average value for the $R$ band in bundled SWNTs. At this stage it
appears as if the vdW interaction with as few as two neighboring
nanotubes in a bundle were sufficient to induce the large pressure
shift observed in bundled SWNTs and attributed to the intertube
interaction.

\paragraph{Elasticity Theory}
The normalized pressure dependence $\omega_0^{-1} \d \omega/\d P \equiv \d
\ln \omega/\d P$ of the radial mode is larger by a factor of $\sim$15 than
that of the tangential modes (section \ref{sec:TM-Raman}) or graphite
\cite{HBS89}. The importance of the vdW forces between the tubes was
illustrated by Thomsen \etal\ \cite{TRGJ99} in the framework of
elasticity theory. Here, a nanotube is approximated by a hollow
cylinder with isotropic elastic properties within the sheet that is
rolled up \cite{Tib84,TRJL99,RTO02a}. Under hydrostatic pressure, the
ratio of the circumferential to axial strain was deduced as
$u_{\Theta\Theta}/u_{zz} = 1.9$. The assumption of similar dependences
on strain, rather than pressure, for the radial and tangential modes,
led Thomsen
\etal\/ to the expectation that the logarithmic pressure dependence of
the $R$ mode should only be about twice as large as that of the
tangential modes in the absence of the vdW forces. Assuming additive
force constants, the logarithmic pressure dependence of the $R$ mode in
a bundle can be separated into a contribution $(1-\alpha)$ from the
radial breathing mode of the isolated tube ($\omega_\text{RBM}$) and a
contribution $\alpha$ from the van-der-Waals interaction
($\omega_\text{vdW}$),
\begin{eqnarray}\label{eq:RBM-elast1}
  \frac{\d \ln \omega}{\d P} & = &
  \left(\frac{\omega_\text{RBM}}{\omega_0}\right)^{\!2}\frac{\d \ln \omega_\text{RBM}}{\d
  P} + \left(\frac{\omega_\text{vdW}}{\omega_0}\right)^{\!2}\frac{\d \ln \omega_\text{vdW}}{\d
  P}\\
  & = &
  (1-\alpha)\frac{\d \ln \omega_\text{RBM}}{\d
  P} + \alpha\frac{\d \ln \omega_\text{vdW}}{\d
  P}  \quad.
\end{eqnarray}
Taking the pressure derivative of the RBM term to be that of the
intramolecular high-energy graphite mode (0.003 GPa$^{-1}$
\cite{HBS89}) and the van-der-Waals component to correspond to the pure
van-der-Waals type $B_{1g}$ mode of graphite (0.15 GPa$^{-1}$
\cite{APZP88}), led to the conclusion that the force constant of the
$R$ mode is to $\alpha = 37\%$ of van-der-Waals nature. This is
equivalent to an up-shift of the radial breathing mode from 136~\WN\
for an isolated tube to 171~\WN\ in a bundle. Although the
van-der-Waals contribution is somewhat overestimated in comparison to
the tight-binding results, it demonstrates its importance and
illustrates how mechanical properties of \emph{mesoscopic}\/ carbon
nanotubes can be studied with the relatively simple continuum
mechanical model \cite{RTO02a}.

\begin{figure}[t]
     \centering
     \includegraphics[width=8cm]{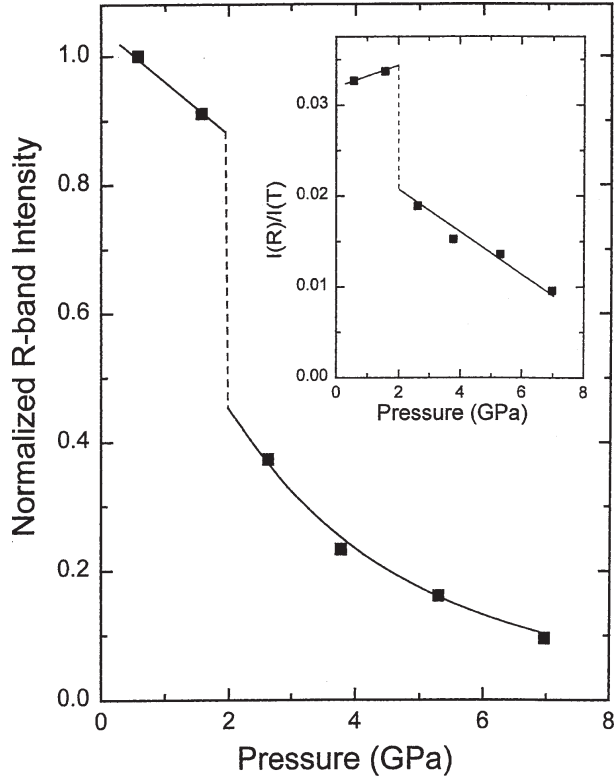}
     \caption{$R$ band integrated intensity of AD-SWNT bucky paper as a
     function of pressure, normalized to its value at 0.5~GPa
     ($\lambda_{ex}=515$~nm). The lines are guides to the eye. The inset
     shows the ratio of the $R$ band to $T$ band integrated intensity. The
     abrupt discontinuity near 2~GPa is apparent in the insert since the
     discontinuity in the $T$ band intensity is not as pronounced as that
     of the $R$ band. Reproduced from Ref.~\cite{VBSR01}.}
     \label{fig:RM-VBSR01-3}
\end{figure}

\paragraph{\boldmath $R$-Mode Disappearance}
Reduction in intensity and broadening of the $R$-mode Raman peak with
increasing pressure as well as its disappearance near 2~GPa have been
confirmed in a number of investigations with excitation at 515~nm
\cite{TRGJ99,TSMS00,PMLK00} and 488~nm \cite{OLSK99}. In one experiment on
arc-discharge SWNT bucky paper \cite{VBSR01}, the $R$ mode could be
followed up to 7~GPa, but a discontinuous drop in intensity was observed
near 2~GPa (Fig.~\ref{fig:RM-VBSR01-3}). The intensity decrease under
pressure was tentatively attributed to a loss in the electronic resonance
of the Raman scattering cross section due to a hexagonal distortion in the
cylindrical cross section of the nanotubes in bundles under compression
\cite{VRRM99}. Further experimental results on the tangential modes under
pressure indicate a subtle structural transition near 2~GPa that will be
discussed in detail in section~\ref{sec:anomaly-2GPa}.

\subsection{Tangential Modes}
\label{sec:TM-Raman}

\begin{figure}[tb]
     \centering
     \includegraphics[height=12cm]{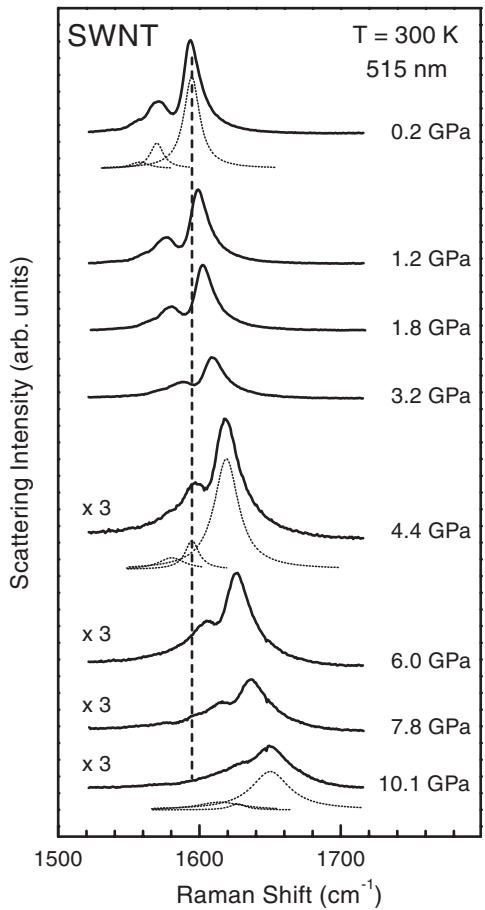}
     \caption{Raman spectra in the tangential-mode region of arc-discharge
     grown SWNTs for increasing hydrostatic pressures up to 10~GPa. Solid
     lines represent the experimental data and dotted ones the
     decomposition of the spectra into three peaks. Reproduced from
     Ref.~\cite{TRJL99}.}
     \label{fig:TM-Specs-TRJL99-1}
\end{figure}

\begin{figure}[tb]
     \centering
     \includegraphics[height=7cm]{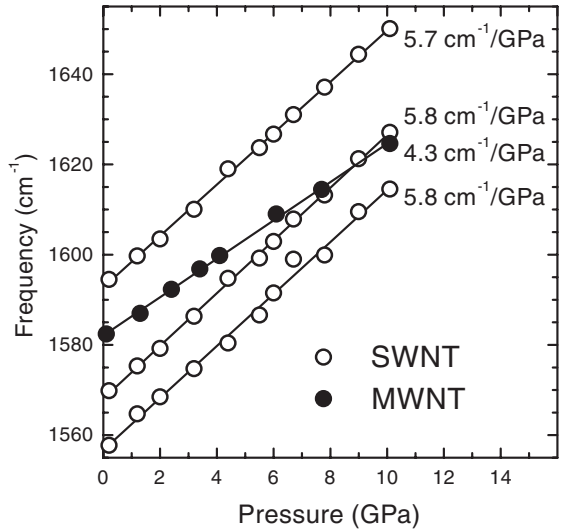}
     \caption{Wavenumbers vs pressure of the three $T$-band components of
     the AD-SWNTs shown in Fig.~\ref{fig:TM-Specs-TRJL99-1} and for the
     $E_{2g}$-like mode of AD multi-wall nanotubes.
     Reproduced from Ref.~\cite{TRJL99}.}
     \label{fig:TM-Specs-TRJL99-3}
\end{figure}

\paragraph{Basic Results}
Figure~\ref{fig:TM-Specs-TRJL99-1} shows Raman spectra of as-prepared
AD-SWNT bundles \cite{TRJL99}. In the region of the $T$ band, the Raman
spectra exhibit three components, all of which shift towards higher
wavenumbers with increasing pressure
(Fig.~\ref{fig:TM-Specs-TRJL99-3}). In this experiment, all components
of the $T$ band exhibited essentially the same pressure dependence of
5.7--5.8~\WN/GPa. Other experiments
\cite{STMS99,TSMS00,VBSR01,TSSK01,RJT00} on AD SWNTs employing
methanol/ethanol as a pressure medium give similar results, i.e.,
wavenumber shifts of 5.3--6.1~\WN/GPa for the most intense component of
the $T$ band (Table~\ref{tab:Raman1}). In PLV-grown samples, the
average pressure-induced shifts appear to be somewhat smaller. All of
the studies on PLV samples \cite{VRRM99,VBSR01,OLSK99,SVRC02p} except
for that by Peters \etal\ \cite{PMLK00} report wavenumber shifts of
4.7--5.1~\WN/GPa. The difference between AD- and PLV-grown samples may
be related to the smaller average diameter of the latter. The
phonon-frequency calculations by Kahn and Lu \cite{KL99b} on bundled
armchair tubes [$(n,n)$ tubes with $n=10\To13$] indicate a correlation
between the pressure dependence of the tangential mode frequencies and
the nanotube diameter. The same applies, however, to the radial mode,
where such a correlation is not recognizable in the available
experimental data.

\paragraph{Structural Transition at 2~GPa}
In their early work, Venkateswaran \etal\ \cite{VRRM99} reported a sudden
drop of the $T$ mode intensity by a factor of 5 between 1.5 and 1.9~GPa
and above 1.9~GPa a broadening of the $T$ band with increasing pressure.
The broadening turned out to be mostly reversible upon pressure release
from the maximum value of 5.2~GPa. In contrast, only about half of the
initial $R$ and $T$ band intensity was recovered, and the peak positions
were shifted to slightly lower wavenumbers by 2--4~\WN. The observation
that the pressure-induced changes were not fully reversible was taken as
an indication that the intertubular contacts within the bundles might
change under pressure.

\begin{figure}[tb]
     \centering
     \includegraphics[width=10cm]{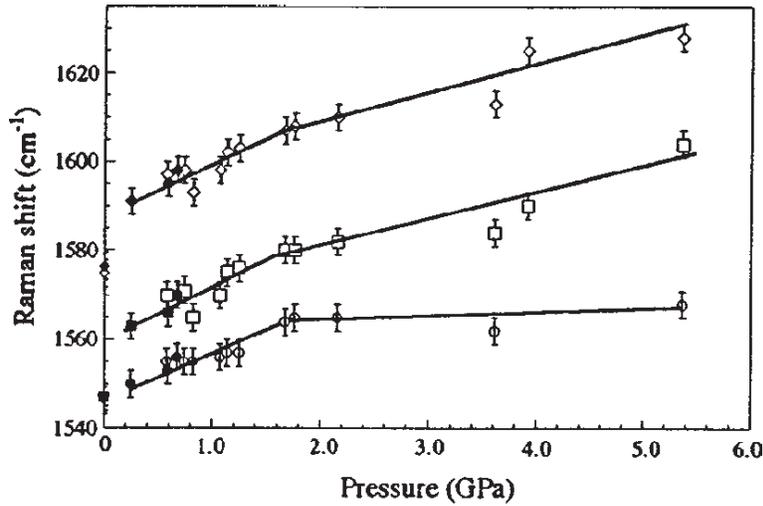}
     \caption{Raman shift vs pressure for tangential modes after Peters
     \etal\ \cite{PMLK00}. The three tangential peaks are represented by
     circles, squares, and diamonds, respectively. Open (closed) symbols
     indicate data taken on pressure increase (decrease).}
     \label{fig:TM-PMLK00-6}
\end{figure}

An abrupt decrease of the pressure dependences of the tangential mode
wavenumbers was first reported by Peters \etal\ \cite{PMLK00}.
Figure~\ref{fig:TM-PMLK00-6} depicts the change in slope of the
tangential modes of PLV-SWNTs at 1.7~GPa. Below the critical pressure,
the modes shift at a rate of $\sim$10~\WN/GPa. Above 1.7~GPa, the
slopes of the two higher-energy components reduce to $\sim$5.9~\WN/GPa
and that of the lower-energy peak becomes as small as 0.7(1)~\WN/GPa.
The peak frequencies were observed to return to their initial values
after pressure release, but only after some delay on the order of 30
minutes. These results motivated a reanalysis of the earlier data of
Venkateswaran \etal\ \cite{VRRM99} that had initially been represented
by a quadratic polynomial. The reanalysis showed that the curvature of
the tangential-band energy shift could possibly be interpreted as a
change in slope of the wavenumber shift near 2~GPa \cite{VBSR01}. The
most intense component of the $T$-band then exhibits a shift of
7.5~\WN/GPa in the range 0--2~GPa and a smaller shift of 4.8~\WN/GPa in
the range 2--5.5~GPa (Table~\ref{tab:Raman1}). These observations were
the first indications of a subtle structural transition in the nanotube
bundles near 2~GPa that will be discussed in detail in
section~\ref{sec:anomaly-2GPa}.

\begin{figure}[t]
     \centering
     \includegraphics[height=8cm]{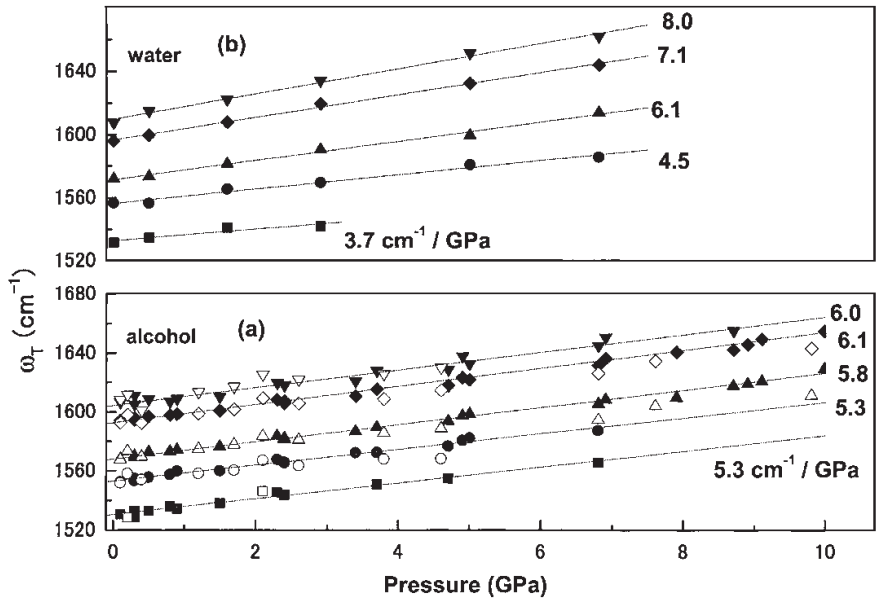}
     \caption{Variation of the tangential mode wavenumbers of AD bucky
     paper with respect to pressure, using (a) a 16:3:1
     methanol/ethanol/water mixture and (b) pure water as pressure media.
     The solid (open) symbols represent data for increasing (decreasing)
     pressures. The straight lines are linear fits to the data with the
     resulting pressure derivatives (in \WN/GPa) given on the figure
     ($\lambda_{ex}=515$~nm).
     Reproduced from Ref.~\cite{TSSK01}.}
     \label{fig:TM-TSSK01-4}
\end{figure}

\paragraph{Shear Strain}
In two of the Raman studies on AD-SWNTs \cite{RJT00,TSSK01}, significant
differences between the pressure shifts of the individual components were
reported. Reich \etal\ \cite{RJT00} interpreted this finding in an
experiment with excitation at 647~nm, i.e., where resonant Raman
scattering mainly from metallic tubes is expected, in terms of shear
strains due the difference in axial and radial compression, as discussed
in the previous section. In armchair tubes, that are always metallic, this
anisotropy is expected to result in different pressure derivatives for the
frequency of modes with axial and circumferential displacement of the
atoms \cite{RJT00}. Teredesai \etal\ \cite{TSSK01}, on the other hand,
conducted the experiment with excitation at 515~nm, where semiconducting
tubes should be in resonance. If a methanol/ethanol/water mixture was used
as pressure medium, the three most intense components shifted at rates of
5.8--6.1~\WN/GPa, whereas two low-energy, low-intensity components of the
$T$ band shifted with 5.3~\WN/GPa (Fig.~\ref{fig:TM-TSSK01-4}). This
difference became much more pronounced, when water was used as pressure
medium. In this case, wavenumber shifts in the range 3.7--8.0~\WN/GPa were
observed. A possible cause for this change are nonhydrostatic stress
components that may arise in water as a pressure medium that solidifies at
1~GPa, whereas the alcohol mixture remains liquid up to 10~GPa.

\begin{figure}[t]
     \centering
     \includegraphics[width=8cm]{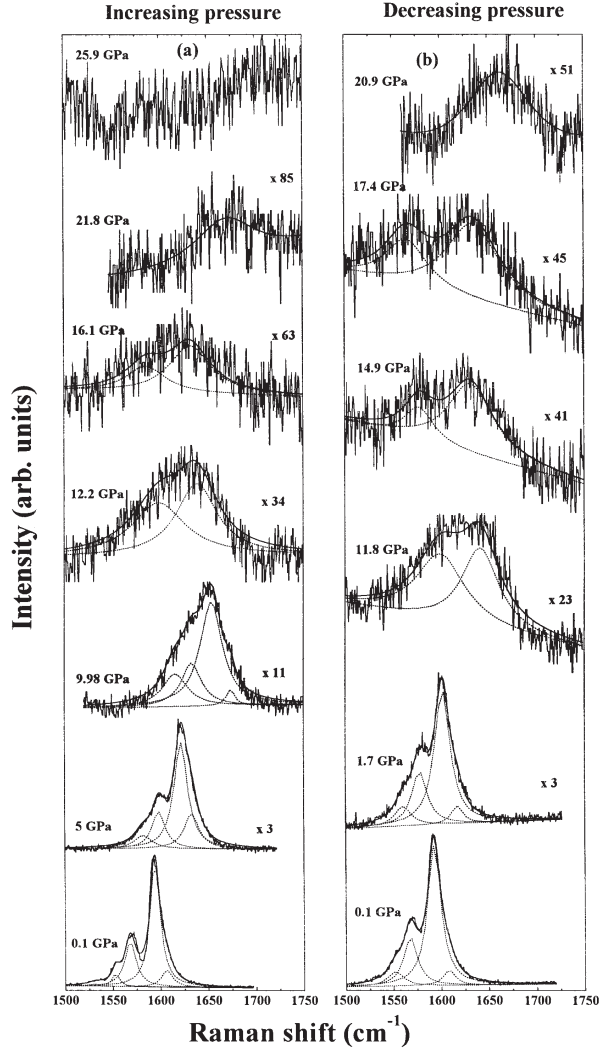}
     \caption{Raman spectra of AD-SWNT bucky paper in the tangential mode region for
     (a) increasing pressures to 26~GPa and (b) decreasing pressures ($\lambda_{ex}=515$~nm).
     Dotted lines represent the decompositions into Lorentzian peaks fitted
     to the data.
     Reproduced from Ref.~\cite{TSMS00}.}
     \label{fig:TM-TSMS00-2}
\end{figure}

\begin{figure}[t]
     \centering
     \includegraphics[width=8cm]{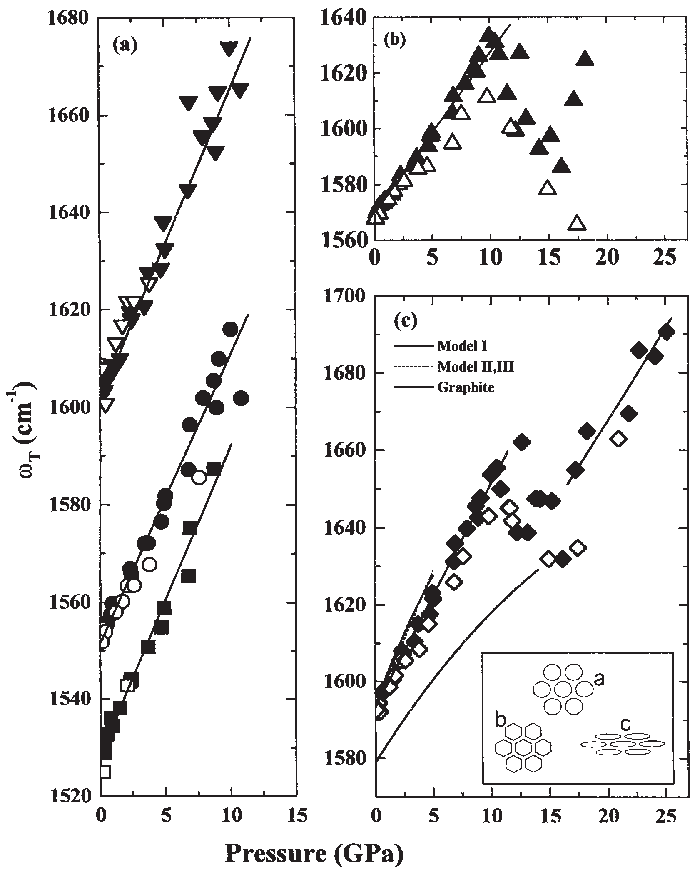}
     \caption{Peak positions as a function of pressure for the Raman
     spectra of AD-SWNT bucky paper shown in Fig.~\ref{fig:TM-TSMS00-2}.
     Peak positions of (a) three components of the $T$ band up to 10~GPa, of (b) the
     second-strongest component (1568-\WN\ at 0~GPa), and of (c) the most
     intense component (1594~\WN\ at 0~GPa). Solid (open) symbols
     represent data for increasing (decreasing) pressures. The straight
     lines at 0--10 GPa are linear fits to the data. Models I--III in panel
     (c) refer to the calculations of Venkateswaran \etal\ \cite{VRRM99}
     in the range 0--5~GPa. Possible structural
     distortions of the nanotube cross section \cite{CNRS99} are sketched in the inset.
     Reproduced from Ref.~\cite{TSMS00}.}
     \label{fig:TM-TSMS00-4}
\end{figure}

\paragraph{Structural Resilience}
The structural stability of SWNTs under compression up to 30~GPa has been
explored by Sood \etal\ \cite{STMS99,TSMS00,TSSK01}.
Figure~\ref{fig:TM-TSMS00-2} depicts Raman spectra recorded on AD-SWNT
bucky paper with a methanol/ethanol/water mixture as the pressure
transmitting medium. The signal intensity decreases continuously with
increasing pressure up to 26~GPa. After pressure release, the initial
Raman spectrum is essentially recovered. Compared to the initial spectrum,
there is only an intensity loss of about 20\% and a moderate increase in
peak width. The peak positions as a function of pressure are shown in
Fig.~\ref{fig:TM-TSMS00-4}. The important result here is a pronounced
softening that occurs for increasing pressures in the range of 10 to
$\sim$16~GPa. Upon pressure decrease, essentially the same anomaly is
observed. It should be mentioned that the anomaly may, in principle, be
related to the solidification of the pressure transmitting medium at
10~GPa and the associated occurrence of shear strains above this pressure.
Experiments with pressure media other than alcohol were, unfortunately,
limited to maximum pressures of 9~GPa (nitrogen \cite{OLSK99}, solid at
2.5~GPa) and 7~GPa (water \cite{TSSK01}, solid at $\sim$1~GPa). Teredesai
\etal\ \cite{TSSK01} argued that the absence of a phase transition up to
6~GPa in the experiment with water would rule out the possibility that the
10-GPa anomaly could be related to the freezing of the pressure medium.
For unambiguous evidence that the anomaly is an intrinsic property of
carbon nanotubes, the softening should, ideally, be reproduced in an
experiment to 15--20~GPa employing a fully hydrostatic pressure medium,
i.e., helium.

In complementary x-ray diffraction work on the same type of sample by
Sharma \etal\ \cite{SKST01}, the $(1\,0)$ reflection of the
two-dimensional triangular lattice of the SWNT bundles vanished at
$\sim$10~GPa. They employed 4:1 methanol/ethanol as a pressure
transmitter. A calculation of diffraction patterns showed that a hexagonal
facetting of the nanotubes does not affect significantly the intensity of
the $(1\,0)$ reflection. Altogether, the observations were interpreted in
terms of a reversible loss of the translational coherence in the nanotube
bundles. It is noteworthy that the $(1\,0)$ reflection was observed to
vanish at a much lower pressure of $\sim$1.5~GPa in x-ray diffraction work
by Tang \etal\ \cite{TQSY00}.


The combined Raman and x-ray diffraction results on the 10-GPa anomaly
were interpreted in terms of a formation of kinks and fins as investigated
in molecular dynamics simulations of highly strained nanotubes by Iijima
\etal\ \cite{IBMB96} and Yakobson \etal\ \cite{YBB96}. The formation of
such distortions would explain the loss of translational coherence. Also,
it would explain the apparent expansion of the triangular lattice that was
observed for increasing pressure just before the disappearance of the
$(1\,0)$ reflection \cite{SKST01}. The lattice expansion may then be the
cause of the Raman mode softening. On the other hand, the axial strain of
the nanotubes present at 10~GPa is much lower than that required to induce
the kinks and fins in the molecular dynamics calculations. Whatever the
structural changes may be, recovery of the initial Raman signature after
pressure release evidences a remarkable resilience of SWNTs under high
pressures.

\subsection{Anomaly at 2~GPa}
\label{sec:anomaly-2GPa}

\paragraph{Summary of Experimental Indications}
In the preceding sections we have seen evidence of discontinuous changes
in the Raman spectra of SWNTs near 2~GPa. In most experiments, the $R$
band vanishes in the pressure range 1.5--3~GPa
(Fig.~\ref{fig:RBM-Specs-VRRM99-1a-3a}). In one study \cite{VBSR01}, the
radial mode could be followed up to 7~GPa, but it exhibited a substantial
drop in intensity at about 2~GPa (Fig.~\ref{fig:RM-VBSR01-3}). Peters
\etal\ \cite{PMLK00} and Venkateswaran \etal\ \cite{VBSR01} reported a
change in slope at 1.7 and 2~GPa, respectively, for the pressure-induced
shift of the $T$ band (Fig.~\ref{fig:TM-PMLK00-6}). In contrast, Teredesai
\etal\ \cite{TSSK01} stated explicitly that they did not observe such
change in slope. Finally, Venkateswaran \etal\ \cite{VRRM99} and
Obraztsova \etal\ \cite{OLSK99} reported broadening of the tangential
peaks to occur only for pressures above 2~GPa.

\begin{figure}[t]
  \centering
  \includegraphics[width=10cm]{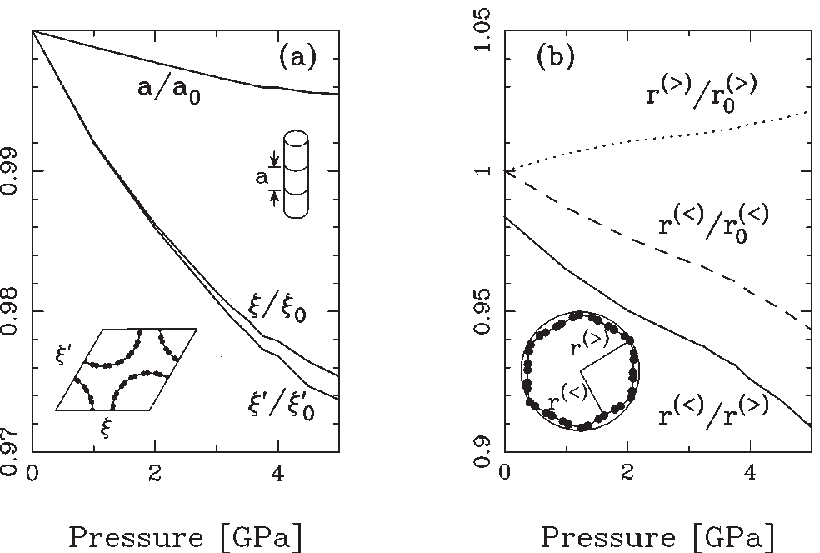}
  \caption{Pressure dependence, as calculated within model I of
      Ref.~\cite{VRRM99}, of (a) the lattice constants and (b) the
      hexagonal distortion of the cross section of an individual $(9,9)$
      nanotube in a bundle. The ratio $r^{(>)}/r^{(<)}$ serves as a measure of
      the distortion. For clarity the distortion depicted in the inset of
      (b) is exaggerated compared to the actual deformation at 5~GPa.
  Reproduced from Ref.~\cite{VRRM99}.}
  \label{fig:2GPa-VRRM99-4}
\end{figure}

\paragraph{Theoretical Studies}
Molecular dynamics calculations (generalized tight-binding method)
\cite{VRRM99} of a $(9,9)$ nanotube bundle with triangular lattice
evidenced a hexagonal distortion of the individual tubes. The facetting
occurred already at zero calculated pressure, and it increased in
magnitude with increasing pressure [Fig.~\ref{fig:2GPa-VRRM99-4}(b)]. The
lattice was allowed to adopt lower than hexagonal, i.e.\
oblique/monoclinic, symmetry. This resulted in a small difference between
the two lattice parameters perpendicular to the nanotubes, which
corresponds to an oval deformation of the tubes
[Fig.~\ref{fig:2GPa-VRRM99-4}(a)]. No structural transition was inferred
from these calculations up to 5~GPa.

\begin{figure}[t]
  \centering
  \includegraphics[width=10cm]{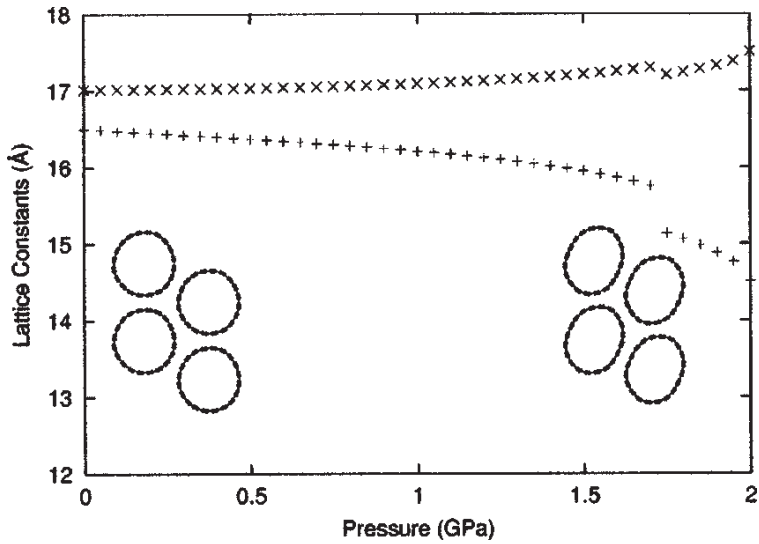}
  \caption{In-plane lattice constants vs pressure for a (10,10) nanorope.
  Insets: Cross sections of tubes at atmospheric pressure and 1.85~GPa.
  Reproduced from Ref.~\cite{PMLK00}.}
  \label{fig:2GPa-PMLK-8}
\end{figure}

Calculations of a $(10,10)$ nanotube rope (universal force field method)
gave rise to a rather different picture \cite{PMLK00}. An important
difference between a triangular lattice of $(9,9)$ and $(10,10)$ tubes is
that the symmetry of the latter tubes is incompatible with the hexagonal
lattice of the bundle. Consequently, already at zero pressure two
different in-plane lattice constants were deduced, the tubes were ovally
distorted, and the lattice of the nanotube bundle was monoclinic rather
than hexagonal (Fig.~\ref{fig:2GPa-PMLK-8}). The calculated distortion
increased continuously with increasing pressure up to 1.75~GPa, where a
sudden transition to a more distorted structure occurred. This transition
was associated with the experimentally observed change in shift rate of
the tangential modes at 1.7~GPa \cite{PMLK00} with the experimental and
theoretical transition pressures being in remarkably good agreement. The
oval distortion of the nanotubes would explain the disappearance of the
radial band at about 2~GPa, because it is no longer an eigenmode.

A subsequent theoretical investigation \cite{SKK02} in the framework of
density-functional theory (DFT) confirmed the different behavior for SWNTs
whose symmetry is compatible or incompatible with a triangular lattice.
For $(10,10)$ nanotubes, a sudden transformation to an oval shape,
accompanied by a transformation of the lattice from near-hexagonal to
pronounced-monoclinic was predicted. If the generalized gradient
approximation (GGA) to the DFT was employed, the transition pressure
amounted to 2.5~GPa and the structure reverted to the near-hexagonal at
0.4~GPa upon pressure release. Using the local density approximation
(LDA), a lower transition pressure of 1.2~GPa was calculated, and the
monoclinic structure persisted down to ambient pressure upon unloading,
i.e., the structure underwent an irreversible change. In contrast, for a
bundle of $(12,12)$ nanotubes a hexagonal deformation was inferred, and
the lattice remained hexagonal over the whole pressure range of the
calculation, 0--6~GPa.

\paragraph{Structural Changes}
Contradicting conclusions were drawn from several diffraction experiments
of SWNTs under pressure. As mentioned before, Tang \etal\ \cite{TQSY00}
reported the disappearance of the $(1\,0)$ reflection of the triangular
lattice under hydrostatic conditions at 1.5~GPa (without, unfortunately,
substantiating this claim by showing the corresponding diffractograms). At
variance with this result, Sharma \etal\ \cite{SKST01} could follow the
$(1\,0)$ reflection up to $\sim$10~GPa, and they did not find evidence for
a structural transition at 1.7~GPa. The latter finding is in accordance
with the fact that Teredesai \etal\ \cite{TSSK01} did not observe, on the
same type of sample, a change in wavenumber shift of the tangential modes
near 1.7~GPa. Likewise, there is no consensus on the question of
reversibility of the structural changes. The results range from persistent
deformations observed in high-resolution transmission electron microscopy
(HRTEM) images after hydrostatic pressurization to only 1~GPa
\cite{Obr01}, over irreversible deterioration of the bundle structure
after applying pressures of more than 4--5~GPa \cite{TQSY00,RGAS01}, to
almost complete reversibility of the Raman spectrum of SWNTs even after
subjecting them to a pressure of 30~GPa under nonhydrostatic conditions
\cite{TSSK01}.

\paragraph{Conclusions}
Given the overall agreement on (i) the discontinuous reduction in
intensity of Raman bands near 2~GPa and (ii) the theoretical results
regarding the oval and hexagonal distortions of bundled nanotubes, one
may interpret the 2-GPa anomaly as follows. In a typical carbon
nanotube sample, a mixture of SWNTs is present with a range of
diameters and chiralities. Part of them may have structures compatible
with the triangular lattice, but for a substantial fraction this will
not be the case. Quite likely, the \emph{abrupt} intensity loss of the
$R$ band is therefore caused by an oval deformation of the nanotubes.
If the nanotube sample contains a large fraction of tubes exhibiting
the ovalization, it may also affect the tangential modes. Otherwise,
the competition between hexagonal and oval distortion may mask or smear
the change in slope so that it becomes difficult to detect. The
\emph{continuous} reduction in intensity may, at least in part,
originate from the strengthened intertube interaction in the bundles
under pressure. It quenches the nearly one-dimensional character of the
electronic structure and should lead to a broadening of the sharp
features in the electronic density of states and consequently reduce
the resonant enhancement of the Raman signal.

The contradicting results compiled above may largely originate from
differences in sample composition in terms of diameters and chiralities
of the SWNTs. Variances in experimental conditions, in particular with
regard to the laser excitation power, may also be of relevance.
Substantial heating of carbon nanotubes due the incident laser
radiation is known to occur already at moderate laser powers. For
example, heating by several 100~K was reported for multiwall nanotubes
with excitation powers of less than 4~mW and a laser spot diameter at
the sample of 25~\mum \cite{HYTZ98}.

\clearpage

\section{Lattice Dynamics of MWNTs}
\label{sec:MWNT-lattice-dynamics}

\begin{figure}[t]
  \centering
  \includegraphics[height=10cm]{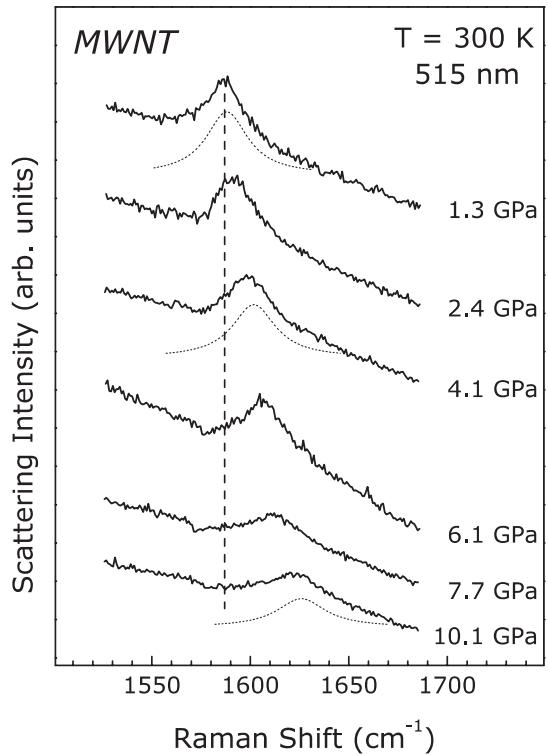}
  \caption{Raman spectra of the high-energy mode of multi-wall carbon
  nanotubes under hydrostatic pressure. Solid lines represent the
  experimental data, dashed lines depict Lorentzian peaks fitted to the
  spectra.
  Reproduced from Ref.~\cite{TRJL99}.}
  \label{fig:MWNT-TRJL99-2}
\end{figure}

\paragraph{Basic Results}
Raman spectra of arc-discharge-grown MWNTs under hydrostatic pressure up
to 10~GPa (methanol/ethanol pressure medium) are depicted in
Fig.~\ref{fig:MWNT-TRJL99-2}. A single symmetric peak was resolved in the
spectral region of the high-energy mode, corresponding to the in-plane
$E_{2g}(2)$ phonons in graphite. With increasing pressure, it shifts
towards higher wavenumbers at a rate of 4.3~\WN/GPa
(Fig.~\ref{fig:TM-Specs-TRJL99-3}). In a second study, MWNTs grown by
chemical vapor deposition (CVD) were studied under pressure with both
methanol/ethanol and condensed helium as a pressure medium. Essentially
the same pressure coefficients were determined for the high-energy mode:
3.8(2)~\WN/GPa and 3.7(1)~\WN/GPa for the experiments with
methanol/ethanol and He, respectively. The difference between the results
of the two studies originates most likely from dissimilar tube dimensions
(inner and outer diameters) in the AD- and the CVD-grown samples. The key
result is that the pressure dependence of the high-energy mode in MWNTs is
substantially smaller than in SWNTs ($\sim$5--6~\WN/GPa, cf.\
Table~\ref{tab:Raman1}) and even smaller than in graphite (4.7~\WN/GPa
\cite{HBS89}).

\paragraph{Elasticity Theory}
The difference between the pressure coefficients of the high-energy modes
in single- and multi-wall nanotubes was discussed in the framework of
elasticity theory \cite{TRJL99}. As mentioned above, this model
approximates the nanotubes by a hollow cylinder with isotropic elastic
properties within the sheet that is rolled up \cite{Tib84,TRJL99,RTO02a}.
The axial strain $u_{zz}$ of a closed hollow cylinder can be derived as
$u_{zz} = -PA(1-2\nu)/E$, where $\nu$ denotes Poisson's ratio, $E$ is
Young's modulus, and $A=R_o^2/(R_o^2-R_i^2)$ is the ratio of total lid
area to the area supported by the cylinder wall. $R_i$ and $R_o$ refer to
the inner and outer diameters of the cylinder, respectively. On the
assumption that the ratio $\rho$ of the pressure coefficients of the
high-energy modes in single- (SW) and multi-wall (MW) tubes equals the
corresponding ratio of axial strain, Thomsen \etal\ \cite{TRJL99} obtained
\begin{equation}\label{eq:MWNT-ratio}
  \rho = \frac{\d \ln \omega^\text{SW} / \d P}{\d \ln \omega^\text{MW} / \d P} =
  \frac{u_{zz}^\text{SW}}{u_{zz}^\text{MW}} =
  \frac{A^\text{SW}}{A^\text{MW}} \approx 1.5 \ ,
\end{equation}
where the inner and outer diameters of the SWNT were chosen as $R_i =
5.2$~Å, $R_o = 8.6$~Å (corresponding to a wall thickness of 3.4~Å), and
for the MWNT the values $R_i = 20$~Å, $R_o = 75$~Å were used. The same
result is obtained if one considers the ratio of circumferential strains.
The result is in fair agreement with their experimental value of $\rho =
1.35$. The difference in pressure derivatives of the high-energy modes in
single- and multi-wall nanotubes can thus be associated with the
differences in wall thickness and tube diameter. The merit of this
approach is not so much to make accurate quantitative predictions, but
rather to help in the interpretation of experimental data in a relatively
simple framework.

\clearpage

\section{Structural Transformations of Carbon Nanotubes}
\label{sec:structural-transformations}

\paragraph{Background} The prospect of transforming graphite or other carbon-rich
matter into diamond has fascinated researchers for a long time
\cite{Haz99}. The most effective approach known to reach this goal is to
subject the starting material (usually graphite) simultaneously to high
pressures and high temperatures \cite{Haz99,Ton92a:C}. The addition of
suitable catalysts can significantly reduce the pressure and/or
temperature required to achieve the conversion. In recent years, carbon
nanotubes have also been considered as a starting material for the diamond
synthesis. It was argued that the curvature of the graphene sheets gives
rise to a partial $sp^3$ hybridization that might result in a formation of
covalent $sp^3$ C--C under milder (pressure and temperature) conditions
\cite{KGBZ02}. Such expectations were promoted by the fact that various
polymerized forms of the fullerene C$_{60}$ can be obtained from a
high-pressure/high-temperature treatment \cite{Sun99}. The possibility of
pressure-induced structural transitions and polymerization of SWNTs was
investigated theoretically \cite{Che98,YGKC00}.

\begin{figure}[t]
  \centering
  \includegraphics[width=10cm]{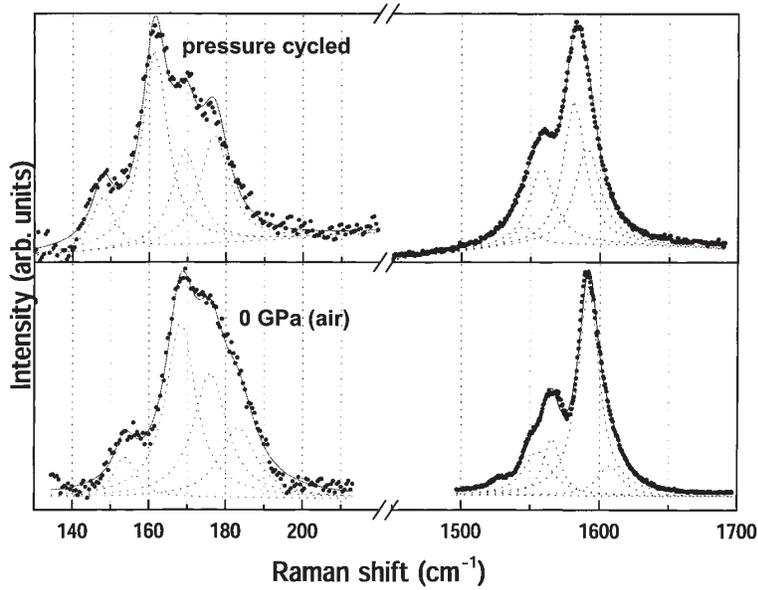}
  \caption{Raman spectra of AD-SWNTs as prepared and after pressure
  release from a maximum quasi-hydrostatic pressure of 26~GPa
  (methanol/ethanol/water pressure medium), showing changes of the $R$
  band and irreversible broadening of the $T$ band. The measured spectra
  are represented by circles, solid lines are fits to the spectra with the
  dotted lines indicating the fitted Lorentzian peaks.
  Reproduced from Ref.~\cite{TSSK01}.}
  \label{fig:transform-TSSK01-6}
\end{figure}

\begin{figure}[t]
  \centering
  \includegraphics[width=10cm]{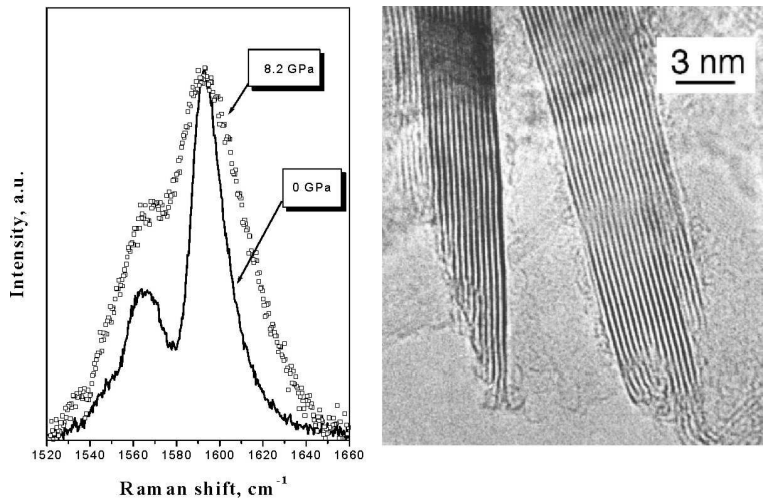}
  \caption{(left) Ambient-pressure Raman spectra of PLV bucky paper as
  prepared and after pressure release from a maximum quasi-hydrostatic
  pressure of 8.2~GPa (nitrogen pressure medium), showing irreversible
  broadening of the $T$ band. (right) HRTEM image of graphitic needles
  that were observed at the base of SWNT bundles after application of
  hydrostatic pressure of 8--9~GPa.
  Reproduced from Ref.~\cite{OLSK99}.}
  \label{fig:transform-OLSK-Fig4+5}
\end{figure}

At ambient temperature and nearly hydrostatic conditions (He pressure
medium), graphite is resistent to a transformation towards cubic
diamond to at least 80~GPa \cite{Gon90}. Teredesai \etal\ \cite{TSMS00}
subjected SWNTs to 26~GPa under quasi-hydrostatic conditions
(methanol/ethanol/water pressure medium) and found the observed changes
to be almost fully reversible when comparing Raman spectra recorded at
0.1~GPa before and after the pressure cycle
(Fig.~\ref{fig:TM-TSMS00-2}). However, after removal of the
pressure-cycled sample from the DAC and evaporation of the pressure
medium, they did observe significant changes of the tangential and
radial bands in the form of a peak down-shift by 6--10~\WN\ and some
broadening (Fig.~\ref{fig:transform-TSSK01-6}). On the basis of X-ray
diffraction measurements on the recovered sample, the authors ruled out
the possibility of graphitization since they did not observe an
increase in intensity of the $(0\,0\,2)$ reflection of graphite. Even
for nonhydrostatic compression (without any pressure medium), changes
to the Raman spectra of SWNTs were reported to be mostly reversible
\cite{TSSK01}. Obraztsova \etal\ \cite{OLSK99} also reported some
persistent broadening of the tangential band after subjecting PLV bucky
paper to a maximum pressure of 8--9~GPa (nitrogen pressure medium). At
variance with the previous authors, however, they did find evidence for
the formation of graphitic needles at the bases of the SWNT bundles in
HRTEM images (Fig.~\ref{fig:transform-OLSK-Fig4+5}). There is a
possibility that graphitic particles have also formed in the
experiments of Teredesai \etal\ but remained undetected in the x-ray
diffractograms due to a rather small particle size (x-ray reflection
broadening due to size effects) and/or small volume fraction compared
to the abundance of $\sim$8\% graphite in the starting material
\cite{SKST01}.

\begin{figure}[t]
  \centering
  \includegraphics[width=10cm]{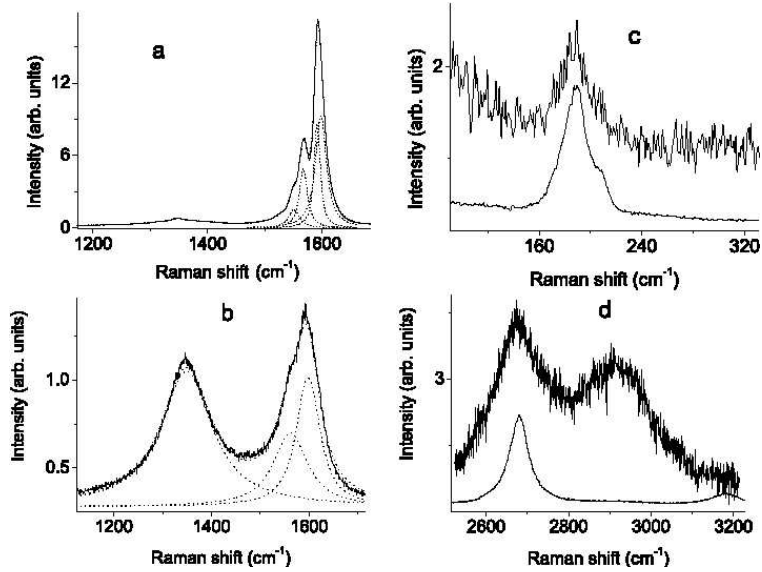}
  \caption{Ambient-pressure Raman spectra in the spectral region of the
        tangential modes ($\sim$1600~\WN) and the defect-related $D$ band
        ($\sim$1350~\WN) of (a) SWNT as grown and (b)
        shear-deformation-treated SWNT after pressure release. (c) $R$
        bands and (d) second-order Raman spectra of as-grown SWNT (lower
        traces) and shear-deformation-treated SWNT (upper traces).
        Reproduced from Ref.~\cite{PKNK02}.}
  \label{fig:transform-PKNK02-3}
\end{figure}

\paragraph{Shear Deformation} Intentional generation of shear strain in a
modified DAC under load was reported to cause a conversion of graphite
into amorphous carbon plus diamond particles (20--500~nm in diameter) at
ambient temperature and pressures as low as 21--25~GPa \cite{ABBD94}. In
the spirit of this approach, Popov \etal\ \cite{PKNK02} subjected PLV
bucky paper to a shear deformation at high pressures. The shear forces
were generated in a DAC by rotating one of the anvils around the axis of
load, with the sample being pressurized without a pressure transmitting
medium. Raman spectra were recorded both \textit{in situ} to monitor the
pressure-induced changes and on the recovered specimen in order to obtain
information on irreversible structural changes
(Fig.~\ref{fig:transform-PKNK02-3}). Apart from some persistent broadening
of the $T$-band, the main effect of the shear-deformation treatment
appears to be a major reduction in intensity of all bands but the one near
1350~\WN\ (defect-related $D$-band). All bands, including the radial mode,
remain observable, though. These observation are consistent with
expectations one may have on the basis of the irreversible changes
reported in previous experiments, e.g., by Obraztsova \etal\ \cite{OLSK99}
and Teredesai \etal\ \cite{TSSK01}.

Popov \etal\ \cite{PKNK02} identified the shear-deformation treated
SWNTs as a ``superhard'' phase of SWNTs. Results of nano-indentation
hardness measurements were presented, in the first place, in order to
substantiate their claim. The recovered material is supposed to have a
hardness in the range 62--150~GPa, i.e., in the hardness range spanned
by cubic boron nitride and diamond. Hardness measurements with
indentations as small as a few tens of nm appear to be a delicate
procedure \cite{RRSH00:ART}, the accuracy of which is difficult to
assess. In the second place, they claim that the shear-deformation
treated SWNT have a bulk modulus of 462--546~GPa, larger than that of
diamond (442--446~GPa) \cite{MAG72,GFDR99}. Besides the fact that a
large bulk modulus is not a sufficient condition for high hardness
\cite{HLB01}, the claim turns out to be untenable in view of the
available experimental data.

A crucial \emph{assumption} in the determination of the bulk modulus (from
the pressure dependence of the $T$ band wavenumber) was that the mode
Grüneisen parameter $\gamma$ is about 1.0 for the $T$-like mode of the
high-pressure treated SWNT. There is, however, no basis for such a
supposition. Due to very anisotropic compression properties, the
$E_{2g}$\mbox{(-like)} modes in graphite and as-grown SWNTs have mode
Grüneisen parameters much smaller than 1, on the order of 0.1
\cite{HBS89}. Therefore, there is no reasonable \emph{a priori} choice for
the Grüneisen parameter, anything in the range 0.1--1 would be reasonable,
and even values outside that range cannot be excluded. The assumption of
$\gamma \approx 1$ does not appear suitable in order to substantiate the
claim of a bulk modulus exceeding that of the most incompressible material
known. So far, a shear-strain-induced transformation of SWNTs into a
material with very high hardness has not been confirmed in independent
investigations.

\begin{figure}[t]
  \centering
  \includegraphics[width=10cm]{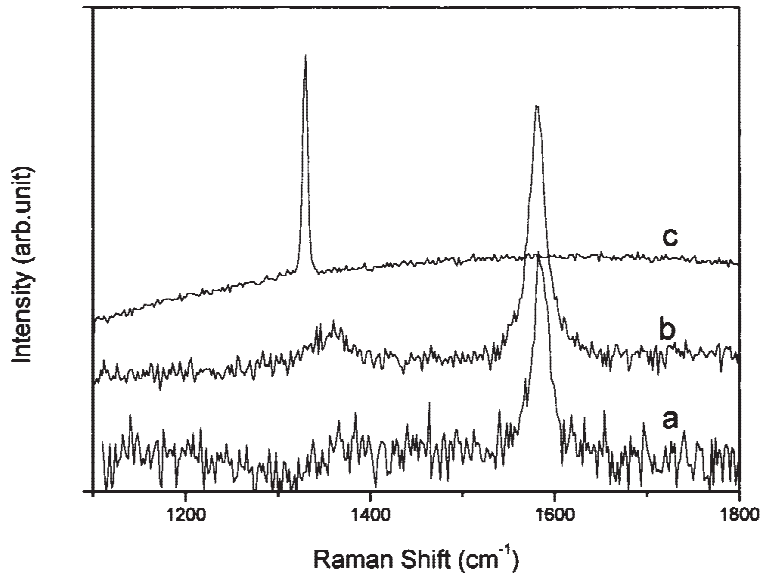}
  \caption{Raman spectra of (a) purified multi-wall carbon nanotubes as
          used as a starting material and after treatment at (b)
          6.0~GPa/1600\degreeC\ and (c) 7.0~GPa/1800\degreeC. The material
          was purified again after the HP/HT treatment \cite{privcomm:Xie03}. Reproduced from
          Ref.~\cite{TCWS00}.}
  \label{fig:transform-TCWS00-2}
\end{figure}

\paragraph{MWNT-to-Diamond Transformation}
Transformation of multi-wall nanotubes into diamond by simultaneous
application of high pressures and high temperatures (HP/HT) was
demonstrated by Tang \etal\ \cite{TCWS00} and Yusa \cite{Yus02}. The
former authors used purified MWNTs (and graphite for comparison) as a
starting material together with a Ni-based alloy as a catalyst. The
sample was pressurized in a boron nitride capsule using a 600-ton
large-volume press. At pressures of 6.0~GPa and 7.0~GPa, the specimen
were heated to 1600\degreeC\ and 1800\degreeC, respectively, for
several seconds. The temperature was then decreased and the pressure
released to ambient conditions in 1~min. The treatment at
6.0~GPa/1600\degreeC\ had little effect on the Raman spectrum of the
MWNTs (Fig.~\ref{fig:transform-TCWS00-2}). After treatment at
7.0~GPa/1800\degreeC\ and purification of the recovered material,
however, the Raman spectrum changed completely. The $T$-band vanished
and a new, relatively narrow peak appeared at 1332~\WN, which coincides
with the position of the Raman peak of diamond. X-ray diffraction data
were reported to support fully the notion of a transformation of the
MWNTs into cubic diamond when subjected to 7.0~GPa and 1800\degreeC\
\cite{TCWS00}. Scanning electron microscopy (SEM) images of the
transformed sample showed that there are many graphite-like flakes
besides the diamond grains. The particles of unconverted carbon and the
catalyst could be removed by etching with \textit{aqua regia} (HCl and
HNO$_{3}$). The SEM images indicated further that the diamond grains
obtained from the nanotubes did not have a crystal form as well-shaped
as those produced from graphite under the same pressure/temperature
conditions. The former had many defects which appeared to be remnants
of nanotubes. Tang \etal\ arrived at the conclusion that multi-wall
carbon nanotubes are not a better source than graphite for the
synthesis of diamond despite the partial $sp^3$-like hybridization.

\begin{figure}[t]
  \centering
  \includegraphics[width=10cm]{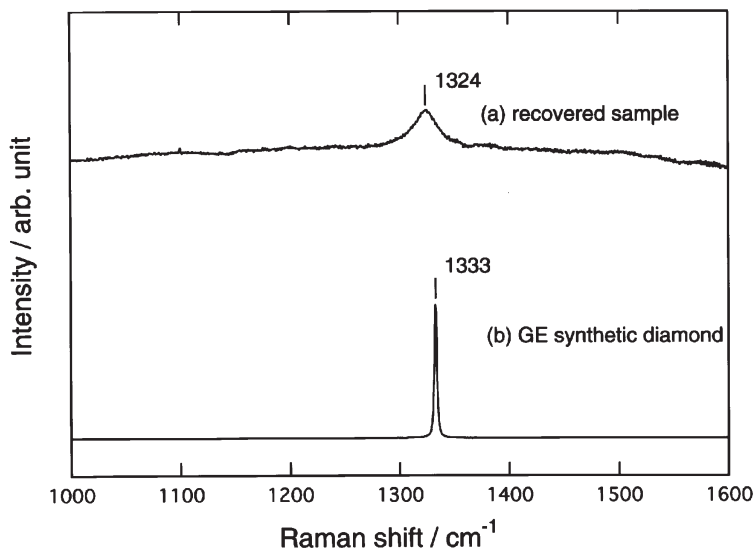}
  \caption{Raman spectra of (a) the transparent material produced from MWNTs
        at $\sim$17~GPa and laser-heating to 2500--3000~K in a DAC and (b)
        a synthetic diamond crystal.
        Reproduced from Ref.~\cite{Yus02}.}
  \label{fig:transform-Yus02-2}
\end{figure}

The possibility of transforming MWNTs into diamond at high pressure in a
laser-heated DAC \emph{without} the use of a catalyst was investigated by
Yusa \cite{Yus02}, in continuation of previous experiments on graphite
\cite{YTMM98}. The starting nanotube sample (5~\mum\ in thickness,
150~\mum\ edge length) was enclosed in KBr to prevent direct thermal
contact between the nanotubes and the diamond anvils. The nanotubes were
heated with a continuous-wave CO$_{2}$ laser (maximum output power of
240~W) to a temperature of 2500--3000~K, as estimated from the emitted
thermal radiation. After a few tens of seconds of heating, a transparent
phase appeared at the irradiated spot. The transparent part of the
recovered sample exhibited a single Raman peak at $\sim$1324~\WN, slightly
below the position of the Raman peak of bulk diamond
(Fig.~\ref{fig:transform-Yus02-2}). Down-shift and broadening of the Raman
peak were attributed to particle size effects \cite{YMMK93}. The view that
the MWNTs were converted to diamond was supported by electron diffraction
on the recovered transparent material. It exhibited only the reflections
of diamond, but not the characteristic $(0\,0\,2)$ reflection of MWNTs and
graphite. Further support came from electron energy loss spectroscopy
(EELS, K-edge absorption and plasmon loss) that gave evidence for $sp^3$
hybridization and loss of the $\pi$-electron signature in the converted
sample. EELS and energy-dispersive x-ray analysis (EDX) did not yield
indication of the presence of any element other than carbon. SEM images
evidenced a granular structure of the recovered material with particle
sizes of less than 50~nm. The upper size limit was related to the diameter
of the initial MWNTs and taken as indication that the nano-sized diamonds
were formed by direct transformation of the nanotubes, without going
through a molten state.

Yusa's observation of a granular structure is quite similar to the earlier
findings of Tang \etal\ \cite{TCWS00}. At variance with the latter
authors, Yusa did not report the presence of graphitic particles. The
assignment of the down-shift of the diamond Raman peak in one of the
experiments to particle size effects may need reconsideration as the
diamond grains appear to have a similar size distribution in both
experiments.

\paragraph{Polymerization of SWNTs}
The possibility to polymerize single-wall carbon nanotubes at high
pressures and temperatures was explored by Khabashesku \etal\
\cite{KGBZ02}. Purified PLV-grown SWNTs, pressed into pellets with a size
of $\sim$2.5~mm, were used as the starting material. Pressures up to
10~GPa were generated in piston-cylinder-type and ``toroid''-type
high-pressure apparatuses. In several runs, samples were heated to
temperatures in the range 200--1500\degreeC\ for 5--300~s, quenched to
ambient conditions, and then characterized by Raman spectroscopy, x-ray
and electron diffraction, electron microscopy, EDX, and in some cases by
electron microprobe analysis (EPMA).

The first set of experiments was performed at quite low a pressure of
1.5~GPa because of the structural changes occurring in SWNTs near 2~GPa
(cf.\ section \ref{sec:anomaly-2GPa}) and because C$_{60}$ starts to
polymerize at such low pressures \cite{Sun99}. Treatment at temperatures
up to 700\degreeC\ mostly affected the $D$ mode near 1300~\WN\ in the
Raman spectrum of the recovered material. This defect-related mode showed
some increase in intensity with increasing maximum temperatures of 300,
500, and 700\degreeC. The $D$ mode was associated here with defects in the
form of intertube $sp^3$-type C--C bonds on the basis of x-ray diffraction
and SEM data. The initial SWNT bundles could mostly be restored after
sonication of the HP/HT-treated material in propanol for two hours.

\begin{figure}[t]
  \centering
  \includegraphics[width=10cm]{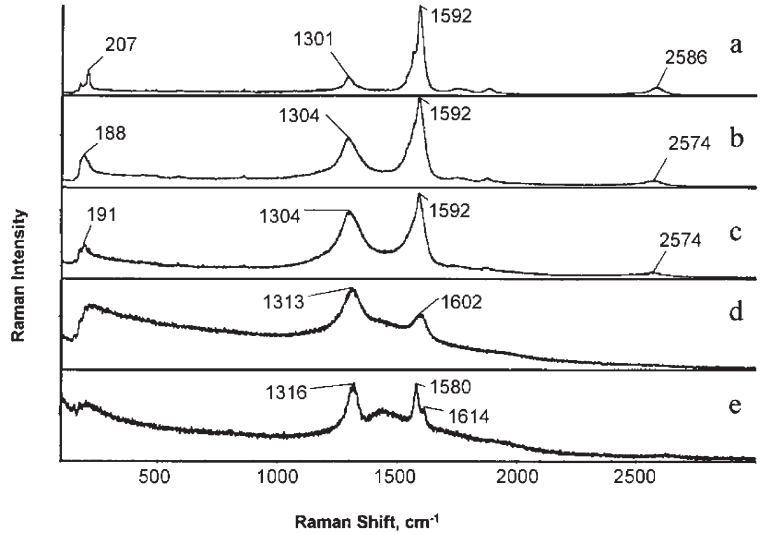}
  \caption{Raman spectra of SWNT samples after HP/HT-treatment at 8.0~GPa
       and temperatures of (a) 200\degreeC, (b) 500\degreeC, (c)
       800\degreeC, (d) 1200\degreeC, and (e) 1500\degreeC\ ($\lambda_{ex}=780$~nm).
       Reproduced from Ref.~\cite{KGBZ02}.}
  \label{fig:transform-KGBZ02-6}
\end{figure}

HP/HT treatment at 8.0~GPa and temperatures up to 1500\degreeC\ lead to
more pronounced changes. For temperatures of 1200 and 1500\degreeC, the
Raman $R$ band near 200~\WN\ and the $D$ band overtone near 2580~\WN\
disappeared, and the $T$ band substantially decreased in intensity
(Fig.~\ref{fig:transform-KGBZ02-6}). The intensity of the $D$ band of all
recovered samples was much larger than in the pristine SWNTs. With
increasing temperature the $D$ band shifted to higher wavenumbers. This
was taken as indication of a possible formation of $sp^3$-type C--C bonds
much like in diamond. Electron microscopy and electron diffraction
indicated a polycrystalline structure of the samples. Treatment at the
highest temperature of 1500\degreeC, however, resulted in a predominant
formation of graphite as evidenced by the Raman and x-ray diffraction
data.

At a pressure of 9.5~GPa, polycrystalline material formed already at
temperatures of 400 and 600\degreeC. Electron diffraction data of the
600\degreeC-sample was taken as evidence for the formation of particles of
hexagonal and cubic diamond. Altogether, these experiments evidenced
irreversible changes to SWNTs when subjected simultaneously to pressures
up to 9.5~GPa and temperatures up to 1500\degreeC. Changes to SWNT bundles
subjected to 1.5~GPa could be reversed by sonication. SWNTs exposed to
9.5~GPa/600\degreeC\ showed signs of a transformation into hexagonal and
cubic diamond.

\paragraph{Conclusions}
A shear-deformation treatment of single-wall carbon nanotubes was reported
to yield a ``superhard'' phase of carbon nanotubes. This claim, however,
appears to require further experimental substantiation. SWNTs treated at
high pressures/temperatures showed indication of the formation of covalent
intertube C--C bonds, i.e., polymerization. At 9.5~GPa and 600\degreeC,
there were signs of a synthesis of hexagonal and cubic diamond particles.
The transformation of multi-wall carbon nanotubes into diamond was
achieved by simultaneous application of high pressures and temperatures.
The results, however, fell short of the expectations that the conversion
might be realized at milder pressure/temperature conditions than required
for the transformation of graphite.

\clearpage

\section{Concluding Remarks}
\label{sec:summary}

Raman scattering studies of carbon nanotubes at high pressures have made
important contributions to our understanding of these materials. In
particular, they have pointed out the importance of the van-der-Waals type
intertubule interaction, that had been assumed for a long time to have
only negligible effect on the structural, vibrational, and electronic
properties. There remain, however, a number of open questions and
challenges. One of the latter is the investigation of truly individual
SWNTs (rather than small bundles) under hydrostatic pressure. This would
address the question of the effect of the van-der-Waals interaction in
very small bundles raised by the investigation of solubilized SWNTs.
Further progress in the preparation of debundled SWNTs may open way to
such experiments. Alternatively, one may consider deposition and
localization of individual nanotubes on the anvil of a DAC in the spirit
of recent experiments \cite{DLBS00} on isolated SWNTs at ambient
conditions.

The interaction of pressure media with carbon nanotubes and possible
intercalation has only been touched upon. It seems that methanol/ethanol
does not enter the interstitial channels in SWNT bundles although
geometrical constraints do not exclude this possibility \cite{VRRM99}.
Helium and nitrogen were employed as pressure media in a few cases
\cite{OLSK99,VBSR01}, but without obvious effect on the nanotube
properties. On the other hand, the condensation of He in the inside and
the interstitial channels of SWNTs was studied theoretically
\cite{GBC00,CCSE00}. Helium was predicted to form a one-dimensional liquid
or solid within the tubes/channels. Intercalation of He or other small
atoms/molecules into carbon nanotubes at high pressure may provide a means
to realize and study quasi-one-dimensional liquids.

Raman spectroscopy has proven a useful tool to monitor attempts to
transform carbon nanotubes into diamond and other ``superhard''
materials. Both single- and multi-wall carbon nanotubes were reported
to transform into diamond at simultaneously high pressures and
temperatures. So far, however, the results fell short of the
expectations that this transformation may occur at milder conditions
than in the case of graphite. Along this line of thought, one may
speculate about the possibility of reacting carbon nanotubes with
nitrogen in a laser-heated DAC in order to synthesize carbon nitride
phases. This class of compounds has attracted a lot of interest in
recent years \cite{HLB01,Mal00,Wan97} because of predictions of a high
bulk modulus of $\beta$-C\sub{3}N\sub{4} comparable to that of diamond
\cite{TH96}. The calculated shear modulus, however, amounted only to
$\sim$60\% of that of diamond so that one should expect a hardness
closer to that of cubic boron nitride. Other carbon nitride phases such
as spinel-type C\sub{3}N\sub{4} \cite{Mal00} may, however, possess a
much higher hardness. The prospect of possibly synthesizing new carbon
phases and substances of high hardness will probably stimulate further
research on the pressure-induced structural transformations of carbon
nanotubes.

\section{Acknowledgements}
I would like to thank U. Venkateswaran, A. K. Sood, U. Schlecht, and K.
Syassen for stimulating and helpful discussions. In addition, I
acknowledge fruitful collaboration with H.~Jantoljak, C.~Thomsen,
U.~Venkateswaran, G.~Duesberg, L. Farina, and K.~Syassen in a number
studies, part of which have been mentioned in this work.

\clearpage \raggedbottom

\begin{small}

\input{Loa_NT-HP-Review.bbl}
\end{small}



\end{document}

%% file: Loa-Raman-table.inc
  \begin{tabular}{lld{2.3}lcccll}
  \toprule
                    & \mc{2}{c}{$R$-band}    & \mc{3}{c}{$T$-band}                \\
    \cmidrule(r){2-3}\cmidrule(l){4-6}
    sample          & \ccol{$\omega_0$} & \ccol{$\d \omega / \d P$} & \ccol{$\omega_0$} & $\d \omega / \d P$ & $P$ range  & $\lambda_{ex}$  & reference & notes\\
                    & \ccol{(\WN)}      & \ccol{(\WN/GPa)}          & \ccol{(\WN)}      & (\WN/GPa)          & (GPa)      & (nm)            &\\

    \midrule
    \emph{experiment}\\
    \addlinespace
    PLV SWNT (1)    & 186        & 7'(1)               & 1593(1)   & $\sim$4.9\tnote{a} & 0--5.5& 515    & \cite{VRRM99,VBSR01}\\
                    &            &                     &           & 7.5           & 0--2       &        & \\
                    &            &                     &           & 4.8           & 2--5.5     &        & \\
    \addlinespace
    PLV SWNT (2)    & 181(3)     & 10'.1(12)           & 1591      & 10.1(11)      & 0--1.7     & 515    & \cite{PMLK00} \\
                    &            &                     &           & 5.8(2)        & 1.7--5.3   &        & \\
    \addlinespace
    PLV bucky paper & 186        & \mbox{$\sim$}10'    & 1592      & 4.9           & 0--10      & 488    & \cite{OLSK99,Obr01} & N$_2$\\
    \addlinespace
    PLV s-SWNT      & 190        & 8'.4                & ?         & $\sim$5.1\tnote{d}  & 0--4.8     & 515    & \cite{VBSR01,SVRC02p} \\
                    &            &                     &           & 7.8           & 0--1.8     &        & \\
                    &            &                     &           & 4.7           & 1.8--4.8   &        & \\
    \addlinespace
    AD SWNT         & 171        & 9'.7(5)             & 1593      & 5.7           & 0--10      & 515    & \cite{TRGJ99,TRJL99} \\
                    &            &                     & 1592      & 6.0           & 0--10      & 647    & \cite{RJT00} \\
    \addlinespace
    AD bucky paper (1)  & 172        & 9'.6                & 1595      & 5.3           & 0--10      & 515    & \cite{STMS99,TSMS00} & M/E/W\\
                    &            &                     &           & 4.5           & 16--25     &        &                          & M/E/W\\
                    &            &                     & 1594      & 6.1           & 0--10      &        & \cite{TSSK01}  & M/E/W \\
                    &            &                     & 1596      & 7.1           & 0--7       &        &                & H$_2$O \\
                    & 170        & 9'.1                &           &               &            &        &                & M/E/W \& H$_2$O\\
                    & 180        & 8'.1                &           &               &            &        &                & M/E/W \& H$_2$O\\
    \addlinespace
    AD bucky paper (2)  & 173        & 8'.3(1)         & 1598(1)   & 5.8(1)        & 0--9       & 515    & \cite{VBSR01} \\
    \addlinespace
    MWNT (1)        &            &                     & 1583      & 4.3           & 0--10      & 515    & \cite{TRJL99} \\
    \addlinespace
    MWNT (2)        &            &                     & 1583(1)   & 3.8(1)        & 0--9       & 515    & \cite{VBSR01} & M/E \\
                    &            &                     &           & 3.7(1)        & 1--9       &        &               & He \\
    \addlinespace
    graphite        &            &                     & 1579      & 4.7           &            &        & \cite{HBS89} \\
    \midrule
    \emph{theory}\\
    \addlinespace
    $(9,9)$ bundle  & 186.2      & \mbox{$\sim$}8'.3\mbox{\tnote{b}}  & 1576 & $\sim$6.6\tnote{c} & 0--5.5 &      & \cite{VRRM99}  & model I\\
    \addlinespace
    $(10,10)$ bundle& 162        & 14                  & 1690      & 11            & 0.1--0.5   &        & \cite{KL99b} \\

  \bottomrule
  \end{tabular}
  \begin{tablenotes}
    \item[a] linear approximation to the relation $\omega(P) = \omega_0 + aP +b P^2$ with $a=7.1(8)$ and $b= -0.4(2)$ given in the original work
    \item[b] linear approximation [$a= 9.6,\  b=-0.65$]
    \item[c] linear approximation [$a= 8.3,\  b=-0.31$]
    \item[d] linear approximation [$a= 7.0,\  b=-0.35$]
  \end{tablenotes}